\documentclass[lettersize,journal]{IEEEtran}
\usepackage{multirow}
\usepackage{bm}
\usepackage{cite} 
\usepackage{amssymb}
\usepackage{graphicx}
\usepackage{subfigure}
\usepackage{algorithm}
\usepackage{algorithmic}
\usepackage{graphicx}
\usepackage{graphics}
\usepackage{textcomp}
\usepackage{subfigure}
\usepackage{tabularx}
\usepackage{times}
\usepackage{multirow}
\usepackage{makecell}
\usepackage{bm}
\usepackage{latexsym}
\usepackage{booktabs}
\usepackage{threeparttable}
\usepackage{epsf}
\usepackage{color,colortbl}
\usepackage{float}
\usepackage[table,xcdraw]{xcolor}
\usepackage{booktabs}
\usepackage{extarrows}
\usepackage{upgreek}

\usepackage{stfloats}  

\begin{document}
	
	\title{Inference-Aware State Reconstruction for Industrial Metaverse under Synchronous/Asynchronous Short-Packet Transmission}
	
	\author{Qinqin Xiong,~\IEEEmembership{Student Member,~IEEE},
              Jie Cao,~\IEEEmembership{Member,~IEEE},
              Xu Zhu,~\IEEEmembership{Senior Member,~IEEE},\\
              Yufei Jiang,~\IEEEmembership{Member,~IEEE},
              and Nikolaos Pappas,~\IEEEmembership{Senior Member,~IEEE}\\

    \thanks{Part of this work has been accepted to be presented at IEEE VTC-Spring 2024 [1].}
    
        }

\maketitle

	\begin{abstract}
	
We consider a real-time state reconstruction system for industrial metaverse. The time-varying physical process states in real space are captured by multiple sensors via wireless links, and then reconstructed in virtual space. In this paper, we use the spatial-temporal correlation of the sensor data of interest to infer the real-time data of the target sensor to reduce the mean squared error (MSE) of reconstruction for industrial metaverse under short-packet transmission (SPT). Both synchronous and asynchronous transmission modes for multiple sensors are considered. It is proved that the average MSE of reconstruction and average block error probability (BLEP) have a positive correlation under inference with synchronous transmission scheme, and they have a negative correlation in some conditions under inference with asynchronous transmission scheme. Also, it is proved that the average MSE of reconstruction with inference can be significantly lower than that without inference, even under weak mean squared spatial correlation (MSSC). In addition, closed-form MSSC thresholds are derived for the superiority regions of the inference with synchronous transmission and inference with asynchronous transmission schemes, respectively. Adaptations of blocklength and time shift of asynchronous transmission are conducted to minimize the average MSE of reconstruction. Simulation results show that the two schemes significantly outperform the no inference case, with an average MSE reduction of more than 50$\%$.

	\end{abstract}
	

\section{Introduction}
Emerging industrial metaverse enables the real-time mapping and operation of physical industrial productions into virtual space, enhancing efficiency \cite{Jagatheesaperumal, Cao1,  Dong1}. For example, the time-varying physical process states in real space are captured by multiple sensors via wireless links and then reconstructed in virtual space by real-time state estimation. The virtual space conducts virtualized operations through a digital platform, \emph{e.g.}, augmented reality (AR), and feeds precise instruction to real space. Accurate real-time state reconstruction is crucial to the effective operation of the industrial metaverse, but challenging due to the time-varying nature of physical processes and the delay and error in transmissions. Mean squared error (MSE) is one of the measures of state reconstruction error \cite{Pappas1, Pappas2, Meng}.

Many data in the industrial metaverse have limited data size (typically 20-250 bytes \cite{3GPP}). Hence, short packet transmission (SPT) is beneficial to reduce transmission delay and MSE of reconstruction \cite{Durisi, Xiong, Polyanskiy}. Unlike the conventional transmission assuming infinite blocklength, SPT suffers from an inevitable block error probability (BLEP) due to limited coding capability, impairing the MSE performance of reconstruction \cite{Polyanskiy}. The data in the industrial metaverse generally exhibit spatial-temporal correlation \cite{Vuran}. For example, the temperature sensor data at key positions of the hot blast stove in the steel smelting. Inference by exploiting the spatial-temporal correlation between data can help to improve the MSE performance of reconstruction under SPT.  In addition, synchronous and asynchronous are typical transmission modes in the industrial metaverse. Synchronous mode transmits multiple sensors’ data simultaneously over different links, enabling a high probability of successful reception at the receiver. Asynchronous mode allows the sensors to separately transmit with a time shift, providing consecutive fresh data at the receiver. Both synchronous and asynchronous transmission modes have the potential to provide a superior MSE performance of reconstruction in inference.

\subsection{Related Work}
The MSE of reconstruction with SPT has been investigated in \cite{Huang, Nadeem, Wu, Roth}. In \cite{Huang}, the average MSE of reconstruction was analyzed for a short-packet linear time-invariant (LTI) system with retransmissions. In \cite{Nadeem}, a non-orthogonal multi-access inspired hybrid automatic repeat request scheme was proposed to reduce the average and packet-level MSE of an LTI system with SPT. In \cite{Wu}, MSE and energy efficient joint optimization was studied under stability constraints. The MSE of reconstruction is related to the information freshness, which is measured by the age of information (AoI), defined by the elapsed time since the generation of the latest successfully received data \cite{Kaul}. In \cite{Roth}, transmission scheduling was studied to reduce MSE of reconstruction and AoI simultaneously. However, the work mentioned above on MSE of reconstruction with SPT ignored spatial-temporal correlation between data and did not consider inference.

Inference is a promising approach to reducing the MSE of reconstruction, where the real-time data of the target sensor with outdated samples is estimated from the fresh samples of the spatially correlated sensors, reducing the AoI used for estimation. The MSE of reconstruction with inference has been widely investigated for distributed estimation systems, where multiple sensors are deployed at different locations to monitor the physical process states of interest \cite{Dong, Rajput, Choi, Cheng, Chen, Zhang, Wattin, Ahmed, Hribar}. In \cite{Dong} and \cite{Rajput}, the average MSE of reconstruction with inference was analyzed under orthogonal and coherent multiple-access channels (MAC), respectively. In \cite{Choi}, a correlation-aware adaptive access method was designed. In \cite{Cheng}, a collaboration compression framework was proposed to reduce the average MSE of reconstruction. In \cite{Chen}, the optimal node number that achieves the best MSE-AoI tradeoff was derived. The work of \cite{Dong, Rajput, Choi, Cheng, Chen} assumed no inference error, \emph{i.e.}, perfect spatial correlation between sensor data, which may not be applicable in practical distributed estimation systems. This was considered in \cite{Zhang, Wattin, Ahmed, Hribar}, where the distance-based spatial correlation model is adopted. In \cite{Zhang}, the optimizations of sensor density and update rate were studied with inference for first-come first-served and last-come first-served service queuing models. In \cite{Wattin}, an AoI-based transmission scheduling scheme was designed to reduce the average MSE of reconstruction. In \cite{Ahmed}, the optimal power allocation was studied to minimize the outage probability of MSE of reconstruction. The work in \cite{Hribar} showed that the inference alongside asynchronous transmission achieves a significant reduction of average MSE of reconstruction over the no inference case, with the assumption of no transmission error and delay.  The SPT error and inference error have an important impact on the MSE performance of reconstruction. However, all the aforementioned work on MSE of reconstruction with inference \cite{Dong, Rajput, Choi, Wattin, Cheng, Chen, Zhang, Ahmed, Hribar} considered no SPT error. The relationship between the MSE of reconstruction and SPT error and inference error has not been analyzed for inference in the previous work \cite{Dong, Rajput, Choi, Wattin, Cheng, Chen, Zhang, Ahmed, Hribar}.

Most existing work on MSE of reconstruction with inference has considered the synchronous transmission mode \cite{Dong, Rajput, Choi, Cheng, Chen, Zhang, Wattin, Ahmed}. Very little work has been conducted to study the ability of asynchronous transmission to assist inference and to enhance the MSE performance of reconstruction. In \cite{Hribar}, a preliminary study for the inference with asynchronous transmission was presented in a two-source distributed estimation system. The quantitative MSE performance comparison between the inference with synchronous transmission and the inference with asynchronous transmission schemes was not provided in \cite{Hribar}, thereby inhibiting the applicable conditions and preferences of the two schemes in the industrial metaverse.  In addition, it is noteworthy that a shorter blocklength leads to a lower AoI at the cost of a higher BLEP, which may impair the MSE performance of reconstruction. Also, the length of time shift in the asynchronous transmission affects the tradeoff between the intra-period AoI and the inter-period AoI. 


In summary, the following open issues on MSE of reconstruction with SPT remain:
\begin{enumerate}
\item What is the relationship between the MSE of reconstruction and SPT error and inference error?
\item What are the preference regions of the inference with synchronous transmission and inference with asynchronous transmission schemes?
\item How to adapt the blocklength and time shift to minimize the average MSE of reconstruction?
\end{enumerate}


\subsection{Contributions}	
Motivated by the above open issues, we investigate inference-aware state reconstruction for industrial metaverse under synchronous and asynchronous SPTs. We reveal important findings that the average MSE of reconstruction and average BLEP are positively correlated under inference with synchronous transmission scheme, and are negatively correlated in some conditions under inference with asynchronous transmission scheme. Also, the average MSE of reconstruction with inference can be significantly lower than that without inference, even under weak spatial correlation related to inference error. The main contributions are as follows.
\begin{itemize}

\item A comprehensive analysis of the relationship between the average MSE of reconstruction and average BLEP and spatial correlation is presented. It is proved that the average MSE of reconstruction is mono-increasing with respect to average BLEP under inference with synchronous transmission scheme. It is decreasing first and then increasing with respect to average BLEP in some conditions under inference with asynchronous transmission scheme. Also, it is proved that the average MSE of reconstruction of both schemes is mono-decreasing with respect to the mean squared spatial correlation (MSSC), defined by the average of the squared spatial correlation between the target sensor and its correlated sensors. Particularly, the average MSE of reconstruction with inference is proved to be significantly lower than that without inference, even under weak MSSC and under weaker than the squared temporal correlation at a transmission period length. 
Furthermore, the upper and lower bounds of the average MSE of reconstruction with respect to the average BLEP and MSSC are proved and derived in closed form.

\item Preference regions of the inference with synchronous transmission and  inference with asynchronous transmission schemes are analyzed, regarding the average MSE performance of reconstruction. Tight approximations of the MSSC thresholds are respectively derived for the superiority regions of the two schemes, which are shown to be mono-decreasing functions of average BLEP and saturate at low average BLEP. Also, closed-form expressions for the average MSE of reconstruction with respect to the MSSC, average received signal-to-noise ratio (SNR) and transmission period are derived. In general, the inference with synchronous transmission scheme is preferable with relatively weak MSSC, low-medium average received SNR and medium transmission period. The inference with asynchronous transmission scheme is desirable in a relatively strong MSSC, low average received SNR and long transmission period regime.

\item The blocklength and time shift are adapted to minimize the closed-form average MSE of reconstruction. The optimal blocklength of the inference with synchronous transmission scheme is derived in closed form. A joint time shift and blocklength optimization (JTSBO) algorithm is proposed for the inference with asynchronous transmission scheme, which significantly outperforms the existing approach with time shift optimization only \cite{Hribar}. The optimization results are shown to be near-optimal by simulation, with a much lower complexity than exhaustive search, and with an average MSE reduction of more than 50$\%$ over the no inference case \cite{Huang, Nadeem, Wu, Roth}.
\end{itemize}

The rest of the paper is organized as follows. Section II presents the system model. Section III presents the analysis of the average MSE of reconstruction for the inference with synchronous transmission and inference with asynchronous transmission schemes, and the analysis of the relationship between the average MSE of reconstruction and average BLEP and spatial correlation. Section IV presents the analysis of the preference regions of the two schemes in terms of the MSSC. Section V is dedicated to the adaptations of blocklength and time shift for the two schemes based on their average MSE analysis in Section III. Simulation results are given in Section VI, followed by the conclusion in Section VII.

\emph{Notations}: Throughout the paper, ${{x}}\sim \mathcal N \left( {{\text{\text 0}},\sigma^{\text 2}} \right)$ denotes that $x$ is a Gaussian random variable with zero mean and variance of $\sigma^{\text 2}$. $Q(u) = \int_u^\infty ({\text {\text 1}}/\sqrt {{\text 2}\pi } ){e^{ - {z^{\text 2}}/{\text 2}}}dz$ is the Q-function. $\mathbb{E}[\cdot]$ denotes expectation. $\frac{{\partial f}}{{\partial x}}$ and $\frac{{{\partial ^{\text2}}f}}{{\partial {x^{\text2}}}}$ denote the first and second derivatives of $f$ with respect to $x$, respectively.

	\begin{figure*}[htbp]  
		\centering
		{\includegraphics[height=4.5cm]{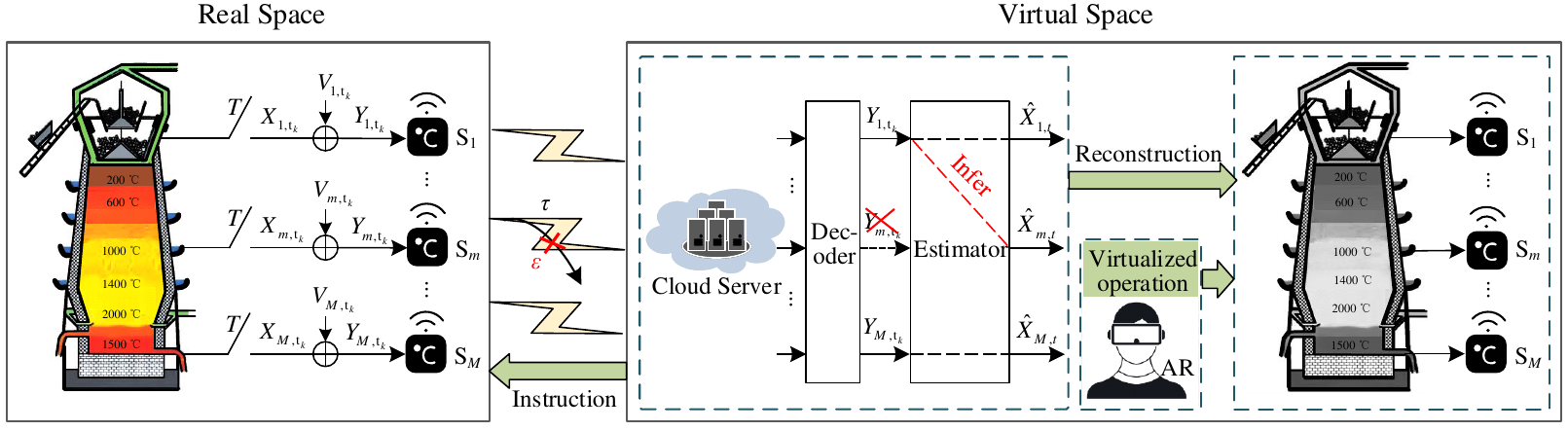}}
		\caption{The inference-aware real-time state reconstruction system model in the industrial metaverse.}\label{fig_1}
	\end{figure*}

\vspace{-5pt}	
\section{System Model}
We consider an inference-aware real-time state reconstruction system with SPT in the industrial metaverse, as shown in Fig.1. $M$ sensors in real space, namely, ${\text S_{\text1}}, {\text S_{\text2}}, \ldots, \text S_M $, are deployed at different locations to monitor the time-varying physical process states of interest, \emph{e.g.}, the temperature of the hot blast stove. They periodically take noisy samples of the physical process states and transmit them to the cloud server in virtual space through independent fading channels. Due to the transmission delay and BLEP, the cloud server immediately reconstructs the real-time data of the physical process with inference after receiving the noisy samples. The digital AR platform conducts virtualized operations on the reconstructed physical process in virtual space and feeds results to the cloud server. Finally, the cloud server calculates precise instructions and transmits them to real space.

\vspace{-10pt}
\subsection{Source Model}
The locations of $M$ sensors are modeled as a homogenous Poisson point process (HPPP) with density $\lambda_{\text d}$ \cite{Zhang}. Time synchronization is assumed among all sensors \cite{Dong, Chen}. The real sample observed by sensor ${\text S}_m$ ($m= {\text 1},{\text 2}, \ldots , M$) at time $t$, denoted by ${X_{m,t}}$, is modeled as a zero mean Gaussian random variable with variance $\sigma _{{\text X},m}^{\text 2}$, \emph{i.e.}, ${{X_{m,t}}} \sim {\mathcal N}({\text 0},\sigma _{{\text X},m}^{\text 2})$. Hence, the real samples of $M$ sensors are joint Gaussian. The spatial-temporal correlation between real samples ${{X_{m,t}}}$ and ${{X_{n,t^\prime}}}$ ($n= {\text 1},{\text 2}, \ldots , M, t^\prime<t$) is defined by 
 \begin{equation}	
{\rho _{mn,t-t^\prime}} = \frac{{ \mathbb{E}[{X_{m,t}}{X_{n,t^\prime}}]}}{ {\sigma _{{\text{X}},m}\sigma _{{\text{X}},n}} } \buildrel \Delta \over = {e^{ - a(t-t^\prime)-br_{mn}}},
 \end{equation}where $r_{mn}$ is the Euclidean distance between sensors ${\text S}_m$ and ${\text S}_n$, with $r_{mm}={\text 0}$ m, $a$ and $b \in \mathbb{R}^ + $ are constant scaling factors, and $e^{ - a(t-t^\prime)}$ and $e^{ - br_{mn}}$ respectively represent the temporal and spatial correlations of the two real samples.  This spatial-temporal correlation model has been widely used in industrial wireless sensor networks \cite{Zhang, Wattin, Ahmed, Hribar, Stein, Pappas3, Pappas4}. The noisy sample observed by sensor ${{\text{S}}_m}$ at time $t$ is denoted by $Y_{m,t}=X_{m,t}+V_{m,t}$, where $V_{m,t} \sim {\mathcal N}({\text 0},\sigma^{\text 2} _{{\text V},m})$ is an independently and identically distributed (i.i.d.) Gaussian observation noise that may include quantization errors \cite{Choi}. The observation SNR of all sensors is assumed to be the same, \emph{i.e.}, $ \frac{{\sigma _{{\text{X,}}m}^{\text 2}}}{{\sigma _{{\text{V,}}m}^{\text 2}}} \buildrel \Delta \over = {\gamma _{\text{o}}}$.

\vspace{-10pt}
\subsection{Transmission Model}
Two transmission modes are considered, \emph{i.e.}, synchronous and asynchronous. For the synchronous transmission mode, $M$ sensors simultaneously transmit packets at the beginning of each transmission period. For the asynchronous transmission mode, $M$ sensors in turn transmit packets with a time shift $h$ in a transmission period. Letting $T$ denote the transmission period, we have $h\le \frac{{T - \tau }}{{M - \text1}}$. Assume that each noisy sample contains $L$ information bits encoded into finite $N$ symbols with symbol duration $T_{\text s}$. Hence, the transmission delay of a packet is given by $\tau=NT_{\text s}$. The channels between sensors and the cloud server are modeled to be i.i.d. quasi-static and block-fading Rayleigh, where the channel gains keep constant during packet transmission. Let $g_m=\sqrt{\beta} h_m$ denote the channel gain between sensor ${\text S}_m$ and the cloud server, where $\beta$ represents the large-scale fading channel gain which includes pathloss and shadowing and is assumed to be the same for all sensors, and $h_m$ represents the small-scale Rayleigh fading with zero mean and unit variance. Let $P_{\text t}$ denote the transmission power at each sensor, $\sigma^{\text 2}_{\text{r}}$ the average AWGN power at cloud server. Then, the instantaneous and average received SNRs at the cloud server are given by ${\gamma _{\text{r},m}} = \beta|h_m|^{\text2} {P_{\text{t}}}/\sigma _{\text{r}}^{\text2}$ and $\bar{\gamma _{\text{r}}} = \beta {P_{\text{t}}}/\sigma _{\text{r}}^{\text2}$, respectively.

For analytical tractability, the normal approximation of BLEP in SPT is employed \cite{Polyanskiy, ZCJ}. The BLEP with an instantaneous received SNR ${\gamma _{\text{r}}}$ can be expressed as $\varepsilon(\gamma_{{\text r}} )  \approx Q\left( {\sqrt {N/V(\gamma_{{\text r}} )} (C(\gamma_{{\text r}} ) - L/N)} \right)$, where $V(\gamma_{{\text r}} ) = {\text {\text 1}} - {({\text {\text 1}} + \gamma_{{\text r}} )^{ - {\text 2}}}$ is the channel dispersion, $C(\gamma_{{\text r}} ) = \ln ({\text {\text 1}} + \gamma_{{\text r}} )$ is the channel capacity per channel use (c.u.), and the subscript $m$ is omitted for notational simplicity. The BLEP can be tightly approximated by transforming the Q-function into the form of segmented linear functions \cite{Yu, Xiong}, \emph{i.e.},
\begin{equation}
\begin{aligned}
 \left\{ \begin{split}
&{\text 1}, \;\;\;\;\;\;\;\;\;\;\;\;\;\;\;\;\;\;\;\;\;\;\;  \gamma_{{\text r}}  < \eta  + {\text 1}/({\text 2}\lambda),\\
&\lambda (\gamma_{{\text r}}  - \eta ) + {\text 1}/{\text 2},\;\, \eta  + {\text 1}/({\text 2}\lambda) \le \gamma_{{\text r}}  \le \eta  -{\text 1}/({\text 2}\lambda),\;\; \\
&{\text 0},\;\;\;\;\;\;\;\;\;\;\;\;\;\;\;\;\;\;\;\;\;\;\;\, \gamma_{{\text r}}  > \eta  - {\text 1}/({\text 2}\lambda),
\end{split} \right.\nonumber
\end{aligned}
 \end{equation}
 \vspace{10pt}where $\eta  = {e^{L/N}} - {\text 1}$ and $\lambda  =  - \sqrt {\frac{N}{{{\text 2}\pi ({e^{{\text 2}L/N}} - {\text 1})}}} $. Then, the average BLEP for Rayleigh fading channel is given by \cite{Yu}
\begin{equation}
\begin{aligned}
\bar \varepsilon  &= \int_{\text 0}^\infty  {\frac{{\text 1}}{{\bar \gamma_{{\text r}} }}{e^{ - \frac{\gamma_{{\text r}} }{{\bar \gamma_{{\text r}} }}}}} \varepsilon (\gamma_{{\text r}} )d\gamma_{{\text r}}\\
 &\approx {\text 1}+ \bar \gamma_{{\text r}} {\lambda}({e^{ - \frac{{\text {\text 1}}}{{\bar \gamma_{{\text r}} }}({\eta} + \frac{{\text {\text 1}}}{{{\text {\text 2}}{\lambda}}})}} - {e^{ - \frac{{\text {\text 1}}}{{\bar \gamma_{{\text r}} }}({\eta} - \frac{{\text {\text 1}}}{{{\text {\text 2}}{\lambda}}})}}).
\end{aligned}
 \end{equation}

\subsection{Performance Metric}
We adopt average MSE of reconstruction as the performance metric. Without loss of generality, we focus on the performance of sensor ${\text S}_m$. The average MSE of reconstruction for sensor ${\text S}_m$ during the time interval $({\text 0},\Gamma]$ is defined as \cite{Ornee}
 \begin{equation} 
{\overline {{\text{MSE}}} _m} = \mathop {\lim }\limits_{\Gamma  \to \infty } \frac{{\text 1}}{\Gamma }\mathbb{E}\left[ {\int_{\text 0}^\Gamma  {{\text{MSE}}_m(t)} dt} \right],
  \end{equation}where ${\text{MSE}}_m(t) = \mathbb{E}[{({X_{m,t}} - {\hat X_{m,t}})^{\text 2}}]$ is the instantaneous MSE of reconstruction and ${\hat X_{m,t}}$ is the reconstructed real-time data of sensor ${\text S}_m$ at the cloud server.

\section{Analysis of Average MSE of Reconstruction for Inference with Synchronous Transmission and Inference with Asynchronous Transmission}

In this section, we derive the closed-form expressions for the average MSE of reconstruction of the inference with synchronous transmission and inference with asynchronous transmission schemes. Based on the results obtained, we analyze the performance gain of inference over no inference, in terms of the average MSE of reconstruction. Also, we analyze the relationship between the average MSE of reconstruction and average BLEP and spatial correlation.

\vspace{-3pt}
\subsection{Average MSE of Reconstruction by Inference with Synchronous Transmission Scheme}

\begin{figure}[t!]
	\centering
	\includegraphics[height=9.6cm]{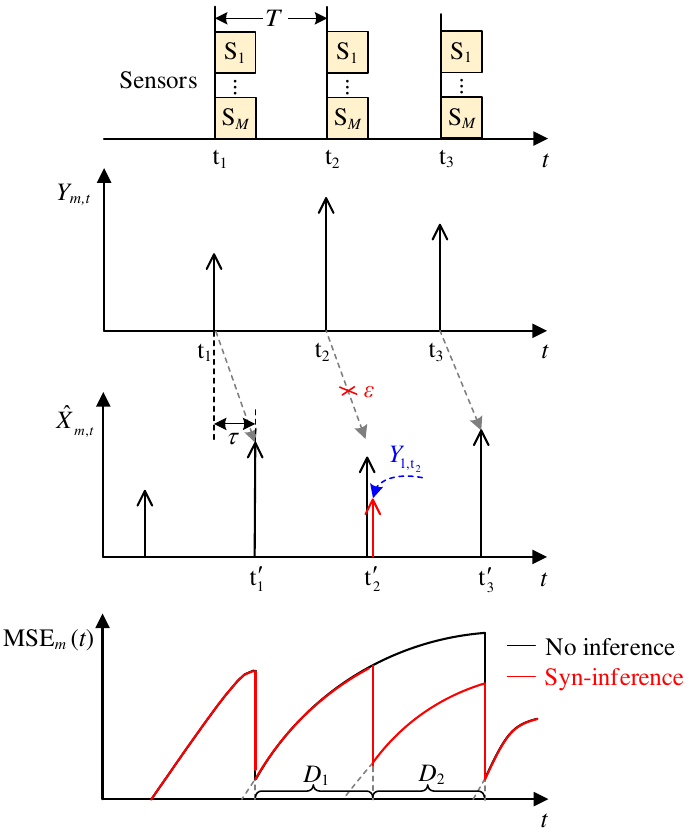}
	\caption{Sampling and reconstruction with and without inference under synchronous transmission mode. The noisy sample $Y_{m,{\text t_{\text2}}}$ of sensor ${\text S}_m$ is lost in the second transmission period. With inference, the cloud server uses the fresh noisy sample $Y_{\text1,{\text t_{\text2}}}$ from sensor ${\text S}_{\text1}$ to reconstruct the real-time data of sensor ${\text S}_m$, which leads to a lower MSE of reconstruction than the case of no inference.}
	\label{fig_syn}
\end{figure}

In the inference with synchronous transmission scheme, all sensors simultaneously transmit noisy samples at the beginning of each transmission period. The cloud server at reception time reconstructs the real-time data of sensor ${\text S}_m$ using the latest successfully received noisy sample that has the strongest spatial correlation to sensor ${\text S}_m$. This enables a reduction of AoI used for estimation in case of transmission error at ${\text S}_m$.

Assume that there are a total of $K$ transmission periods in the time interval (0, $\Gamma$]. Let ${\text t}_k$ denote the transmission start time of the $k$-th ($k = \text1,\text2, \ldots, K $) period. Let ${{\mathbb{Y}}_{u_{\text s}(t)}}$ denote the set of the latest noisy samples that are successfully received by cloud server before time $t$, with $u_{\text s}(t)$ being the generation time. The selected noisy sample used to estimation for sensor $\text{S}_m$ at reception time (${\text t}_k+\tau$) is given by ${Y_{j^*,u_{\text s}(t)}} =\arg \mathop {\max }\limits_{{Y_{j,u_{\text s}(t)}} \in {{{\mathbb{Y}}_{u_{\text s}(t)}}}} {e^{-br_{mj}}}$. With a minimum mean squared error (MMSE) estimator, the reconstructed real-time data of sensor ${\text S}_m$ is given by
  \begin{equation}
\hat X_{m,t}^{{\text{Syn-infer}}} =  \mathbb{E}[{X_{m,t}}|{Y_{j^*,u_{\text s}(t)}}] = \frac{{{\sigma _{{\text{X}},m}}{\gamma _{\text{o}}}{\rho _{mj^*,t-u_{\text s}(t)}{Y_{j^*,u_{\text s}(t)}}}}}{{{\sigma _{{\text{X}},j^*}}({\gamma _{\text{o}}} + {\text 1})}}.
 \end{equation}The instantaneous MSE of reconstruction is given by
   \begin{equation}
\text{MSE}_m^{{\text{Syn-infer}}}(t,j^*) = \sigma _{{\text{X,}}m}^{\text 2}({\text 1} - \frac{{{\gamma _{{\text{o}}}{e^{ - {\text 2}a(t-u_{\text s}(t)) - {\text 2}b{r_{mj^*}}}}}}}{{{\gamma _{\text{o}}} + {\text 1}}}).
   \end{equation}
Assume that there are a total of $V$ successful transmission periods in the time interval of $({\text 0},\Gamma ]$, each of which has at least one noisy sample successfully received by the cloud server. Let ${\text t}_{v}^\prime$ denote the transmission end time of the $v$-th $(v={\text 1},{\text 2},\ldots, V)$ successful transmission period. Then, the time interval between two consecutive successful transmission periods can be expressed as $D_{v}={\text t}_{v+{\text 1}}^\prime-{\text t}_{v}^\prime=zT+T$, where $z (z={\text 0},{\text 1},\ldots)$ is the numbers of failed transmission periods during $D_{v}$. We reindex the sensors according to their spatial correlations to sensor ${\text S}_m$ for ease of calculation. Let $\text{S}_{\tilde s}$ denote the sensor with the $\tilde s$-th $(\tilde s = \text1, \text2, \ldots ,M)$ strongest spatial correlation to sensor ${\text S}_m$, with $\tilde {\text1}=m$. From Fig. {\text {\text 2}}, the average MSE of reconstruction for sensor ${\text S}_m$ by the inference with synchronous transmission scheme can be calculated as
\vspace{-5pt}
\begin{align}
&\overline {{\text{MSE}}} _m^{{\text{Syn-infer}}}\nonumber\\
&= \mathop {\lim }\limits_{V \to \infty } \frac{{\sum\limits_{v = {\text 1}}^{{V}} \mathbb{E}{\left[ {\int_{{{{\text{t}}}_{v}^\prime}}^{{{\text{t}}_{v + {\text 1}}^\prime} } {{\text{MSE}}_m^{{\text{Syn-infer}}}(t,j^*)} dt} \right]} }}{{\sum\limits_{v = {\text 1}}^{{V}} {{D_{v}}} }}\nonumber\\
&= \frac{\mathbb{E}{\left[{\int_\tau ^{\tau + D_v} {\sigma _{{\text X},m}^{\text 2}} {({\text 1} - \frac{{{\gamma _{{\text{o}}}{e^{ - {\text 2}at - {\text 2}b{r_{mj^*}}}}}}}{{{\gamma _{\text{o}}} + {\text 1}}})}dt} \right]}}{{\mathbb{E}[{D_{v}}]}}\nonumber\\
&=\sigma _{{\text{X,}}m}^{\text 2} - \frac{{\sigma _{{\text{X,}}m}^{\text 2}{\gamma _{\text{o}}}{e^{ - {\text 2}a\tau }}\mathbb{E}[{e^{ - {\text 2}b{r_{m{j^*}}}}}]({{\text 1} - \mathbb{E}[{e^{ - {\text 2}a{D_v}}}]})}}{{{\text 2}a({\gamma _{\text{o}}} + {\text 1}){{\mathbb{E}[{D_v}]}}}}\nonumber\\
& = \sigma _{{\text{X,}}m}^{\text 2} - \frac{\sigma _{{\text{X,}}m}^{\text 2}{{\gamma _{\text{o}}}{e^{ - {\text 2}a\tau }}
({\text 1} - {e^{ - {\text 2}aT}})({\text 1} - \bar \varepsilon ) \sum\limits_{\tilde s = {\text 1}}^{M} {{e^{ - {\text 2}b{r_{m\tilde s}}}}{{\bar \varepsilon }^{\tilde s-\text1}}}}}{{{\text 2}aT({\gamma _{\text{o}}} + {\text 1})({\text 1} - {e^{ - {\text 2}aT}}{{\bar \varepsilon }^M})}},
\label{mse_syn}
\end{align}where the expectations of ${e^{ - {\text 2}b{r_{m{j^*}}}}}$, $D_{v}$ and $e^{-\text2aD_{v}}$ are given by $\mathbb{E}[{e^{ - {\text 2}b{r_{m{j^*}}}}}]=\sum\limits_{\tilde s = {\text{1}}}^{M} {\frac{{e^{ - {\text{2}}b{r_{m\tilde s}}}}{{{\bar \varepsilon }^{\tilde s-\text1}}(\text1 - \bar \varepsilon )}}{{\text1 - {{\bar \varepsilon }^M}}}} $, $\mathbb{E}[{D_v}] = \sum\limits_{z = {\text{0}}}^\infty  (zT + T) $ ${{\bar \varepsilon }^{Mz}}({\text{1}} - {{\bar \varepsilon }^M}) = \frac{T}{{{\text{1}} - {{\bar \varepsilon }^M}}}$ and $\mathbb{E}[{e^{ - \text2a{D_v}}}]= \sum\limits_{z =\text0}^\infty  {{e^{ - \text2a(zT + T)}}} {{\bar \varepsilon }^{Mz}}$ $(1 - {{\bar \varepsilon }^M}) = \frac{{{e^{ - \text2aT}}(\text1 - {{\bar \varepsilon }^M})}}{{\text1 - {e^{ - \text2aT}}{{\bar \varepsilon }^M}}}$, respectively. The complex form of  the relationship between the $\overline {\text{MSE}} _{m}^{{\text{Syn-infer}}}$ and spatial correlation factor ${{e^{ - b{r_{mn,n \ne m}}}}}$ hinders further analysis. To this end, define the MSSC in terms of sensor $\text S_m$ as $\bar \rho _m^{{\text{spatial}}} = \frac{\text1}{{M - \text1}}\sum\limits_{n=\text 1,n\ne m}^{M} {{e^{ - \text2b{r_{mn}}}}}$. An approximation of $\overline {{\text{MSE}}} _m^{{\text{Syn-infer}}}$ is given by
\begin{align}
\overline {{\text{MSE}}} _{m,{\text{appr}}}^{{\text{Syn-infer}}} = \overline {{\text{MSE}}} _m^{{\text{Syn-infer}}}\left| {_{{e^{ - \text2b{r_{m\tilde s}}}} = \bar \rho _m^{{\text{spatial}}},\tilde s > \text1}}, \right.
\label{mse_syn_appr}
\end{align}whose tightness is verified by Fig. \ref{fig_MSE_N} in the simulation results. 
For comparison, by setting $M=$ 1 in (\ref{mse_syn}), the average MSE of reconstruction  for sensor ${\text S}_m$ by no inference is given by
\begin{align}
\overline {{\text{MSE}}} _m^{{\text{No infer}}}= \sigma _{{\text X},m}^{\text 2} - \frac{\sigma _{{\text X},m}^{\text 2}{{\gamma _{\text o}}{e^{ - {\text 2}a\tau }}({\text 1} - {e^{ - {\text 2}aT}})({\text 1} - {{\bar \varepsilon }})}}{{{\text 2}aT({\gamma _{\text o}} + {\text 1})({\text 1} - {e^{ - {\text 2}aT}}{{\bar \varepsilon }})}}.
\label{mse_noinfer}
\end{align}From (\ref{mse_syn_appr}) and (\ref{mse_noinfer}), compared to no inference, the inference achieves the performance gain of average MSE of reconstruction as ${G^{{\text{Infer}}}} = \frac{{\overline {{\text{MSE}}} _m^{{\text{No infer}}}}}{{\overline {{\text{MSE}}} _m^{{\text{Syn-infer}}}}} \approx \frac{{\overline {{\text{MSE}}} _m^{{\text{No infer}}}}}{{\overline {{\text{MSE}}} _{m,{\text{appr}}}^{{\text{Syn-infer}}}}} \buildrel \Delta \over = {G_{{\text{appr}}}^{{\text{Infer}}}}$, which is larger than one with $\bar \rho _m^{{\text{spatial}}} > \frac{{{e^{ - {\text{2}}aT}}(\text1 - \bar \varepsilon )}}{{{\text{1}} - {e^{ - {\text{2}}aT}}\bar \varepsilon }}$. Based on this result, we provide Proposition 1.

\textbf{Proposition 1.} Given average BLEP $\bar \varepsilon$ and the squared temporal correlation ${{e^{ - {\text{2}}aT}}}$ of data at a transmission period length $T$, the inference can achieve a lower average MSE of reconstruction than the no inference case as long as the MSSC $\bar \rho _m^{{\text{spatial}}} > \frac{{{e^{ - {\text{2}}aT}}(\text1 - \bar \varepsilon )}}{{{\text{1}} - {e^{ - {\text{2}}aT}}\bar \varepsilon }} \buildrel \Delta \over = \bar\rho _{m,{\text{thr1}}}^{{\text{spatial}}} $.

The MSSC threshold $\bar\rho _{m,{\text{thr1}}}^{{\text{spatial}}}$ increases with a lower average BLEP and saturates to $e^{-\text2aT}$ at low average BLEP, \emph{i.e.}, $\rho _{m,{\text{thr1}}}^{{\text{spatial}}}\le e^{-\text2aT}$. With this observation, we derive Lemma 1.

\textbf{Lemma 1.} The inference can achieve a lower average MSE of reconstruction than the no inference case, even under weak MSSC and under weaker than the squared temporal correlation at a transmission period length,  \emph{i.e.}, $\bar \rho _m^{{\text{spatial}}}<e^{ - {\text{2}}aT}$. When the MSSC $\bar \rho _m^{{\text{spatial}}}\ge e^{ - {\text{2}}aT}$, the average MSE reduction achieved by inference is independent of average BLEP.

Lemma 1 implies that it is easy to obtain the performance gain of average MSE of reconstruction from inference. For example, when the squared temporal correlation ${{e^{ - {\text{2}}aT}}}=$ 0.5 and average BLEP $\bar \varepsilon=$ 0.3, the minimum allowable MSSC is $\bar\rho _{m,{\text{thr1}}}^{{\text{spatial}}}= \frac{{{e^{ - {\text{2}}aT}}(\text1 - \bar \varepsilon )}}{{{\text{1}} - {e^{ - {\text{2}}aT}}\bar \varepsilon }}=$ 0.4, which is smaller than ${{e^{ - {\text{2}}aT}}}$. The weak MSSC causes large inference error, which is supposed to increase the MSE. Also, one may intuitively think that the spatial correlation weaker than the temporal correlation cannot enable an MSE reduction. However, we derive a surprising result that the weak MSSC and weaker than the squared temporal correlation still allows the average MSE of reconstruction lower than the no inference case. The inference significantly reduces the AoI used for estimation in case of transmission error, especially under long transmission period and high BLEP, leading to a significant tolerance of spatial correlation.

\vspace{-10pt}
\subsection{Average MSE of Reconstruction by Inference with Asynchronous Transmission Scheme}

In the inference with asynchronous transmission scheme, $M$ sensors in turn transmit noisy samples with a time shift $h$ during a transmission period. The cloud server at the reception time of every sensor reconstructs the real-time data of sensor ${\text S}_m$ with inference. This provides consecutive fresh noisy samples at the cloud server and a lower AoI than the inference with synchronous transmission scheme.

	\begin{figure}[!]
	\centering
	\includegraphics[height=9.6cm]{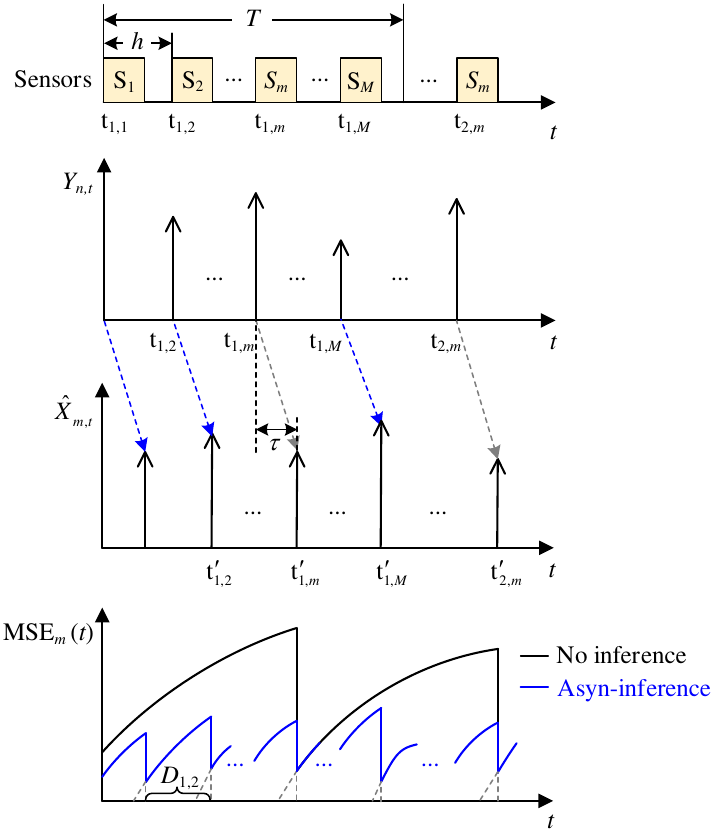}
	\caption{Sampling and reconstruction with and without inference under asynchronous transmission mode. With inference, the cloud server at the reception time of every sensor uses the latest successfully received noisy sample to reconstruct the real-time data of sensor $\text S_m$, which leads to a lower MSE of reconstruction than the case of no inference.}
	\label{fig_asyn}
\end{figure}

Let $ {\text t}_{k,n}$ denote the transmission start time of sensor ${\text S}_n$ in the $k$-th transmission period, ${Y_{l,u_{\text{as}}(t)}}$ the latest successfully received noisy sample before time $t$, with $u_{\text{as}}(t)$ being the generation time, under asynchronous transmission mode. The reconstructed real-time data of sensor ${\text S}_m$ at reception time (${\text t}_{k,n}+\tau$) is given by
\begin{equation}
\begin{aligned}
\hat X_{m,t}^{{\text{Asyn-infer}}} = [{X_{m,t}}|{Y_{l,u_{\text{as}}(t)}}] = \frac{{{\sigma _{{\text{X}},m}}{\gamma _{{\text{o}}}}{\rho _{ml,t-u_{\text{as}}(t)}}}}{{{\sigma _{{\text{X}},l}}({\gamma _{{\text{o}}}} + {\text 1})}}{Y_{l,u_{\text{as}}(t)}}.
\end{aligned}
\end{equation}The instantaneous MSE of reconstruction is given by
\begin{equation}
\begin{aligned}
{\text{MSE}}_m^{{\text{Asyn-infer}}}(t,{l}) = \sigma _{{\text{X}},m}^{\text 2}({\text 1} - \frac{{{\gamma _{\text{o}}}{e^{ - {\text 2}a(t-u_{\text{as}} (t) )- {\text 2}b{r_{m{l}}}}}}}{{{\gamma _{\text{o}}} + {\text 1}}}).
\end{aligned}
\end{equation}Assume that $G_v$ noisy samples are successfully received by the cloud server in the $v$-th successful transmission period. Let ${  {\text t}^\prime}_{v,g}$ denote the reception time of the $g$-th $(g = {\text 1},{\text 2}, \ldots ,{G_v})$ successful noisy sample in the $v$-th successful transmission period. The inter-reception time of asynchronous transmission can be expressed as $D_{v,g}={{{{{\text t}^\prime}}_{v,g}}}-{{{{{\text t}^\prime}}_{v,g-{\text 1}}}}$, with ${{{{{\text t}^\prime}}_{v,{\text 0}}}}={{{{{\text t}^\prime}}_{v-{\text 1},G_{v-{\text 1}}}}}$ and ${D_v} = \sum\nolimits_{g = {\text 1}}^{{G_v}} {{D_{v,g}}} $. From Fig. 3, the average MSE of reconstruction for sensor ${\text S}_m$ by inference with asynchronous transmission scheme can be derived as
\begin{align}
&\overline {\text {MSE}} _m^{{\text{Asyn-infer}}}\nonumber\\
& = \mathop {\lim }\limits_{V \to \infty } \frac{{\sum\limits_{v = {\text {\text 1}}}^V {\left[ {\int_{{{{{ {\text t}}}}^\prime_{v,{G_{v - {\text {\text 1}}}}}}}^{{{{{ {\text t}}}}^\prime_{v,{G_v}}}} {{\text{MSE}}_m^{{\text{Asyn-infer}}}(t)dt} } \right]} }}{{\sum\limits_{v = {\text {\text 1}}}^V {{D_v}} }}\nonumber\\
&=\frac{{\mathbb{E}\left[ {\int_{{{{{{\text t}}}}^\prime_{v,g - {\text {\text 1}}}}}^{{{{{ {\text t}}}}^\prime_{v,g}}} {{\text{MSE}}_m^{{\text{Asyn-infer}}}(t)dt} } \right]}}{{\mathbb{E}[{D_{v,g}}]}}\nonumber\\
&=\frac{{\mathbb{E}\left[ {\int_\tau ^{\tau  + {D_{v,g}}} {\sigma _{{\text{X}},m}^{\text 2}({\text {\text 1}} - \frac{{{\gamma _{{\text{o}}}{e^{ - {\text 2}at - {\text 2}b{r_{m{l}}}}}}}}{{{\gamma _{{\text{o}}}} + {\text {\text 1}}}})dt} } \right]}}{{\mathbb{E}[{D_{v,g}}]}}\nonumber\\
&=\sigma _{{\text{X}},m}^{\text 2} - \frac{{\sigma _{{\text{X}},m}^{\text 2}{\gamma _{\text{o}}}{e^{ - {\text 2}a\tau }\mathbb{E}[{e^{ - {\text 2}b{r_{m{l}}}}}({\text {\text 1}} - {e^{ - {\text 2}a{D_{v,g}}}})]}}}{{{\text 2}a({\gamma _{\text{o}}} + {\text {\text 1}}){\mathbb{E}[{D_{v,g}}]}}}\nonumber\\
&=\sigma _{{\text{X}},m}^{\text 2} - \frac{{\sigma _{{\text{X}},m}^{\text 2}{\gamma _{\text{o}}}{e^{ - {\text 2}a\tau }}({\text 1} - \bar \varepsilon )\sum\limits_{n = {\text{1}}}^M {{e^{ - {\text{2}}b{r_{mn}}}}\Psi_n}}}{{{\text 2}aT({\gamma _{\text{o}}} + {\text 1})({\text 1} - {e^{ - {\text 2}ah}}\bar \varepsilon )}},
\label{mse_asyn}
\end{align}where $\Psi_n =\text1 -  {e^{-{\text{2}}ah}} + \frac{{{{e^{-{\text{2}}ah}} {\bar \varepsilon }^M}({\text{1}} - \bar \varepsilon )({e^{ - {\text{2}}ahM}} - {e^{ - {\text{2}}aT}})}}{{{{(\bar \varepsilon {e^{ - \text2ah}})}^n}{\text{(1}} - {e^{ - {\text{2}}aT}}{{\bar \varepsilon }^M})}}$. The expectations of $D_{v,g}$ and ${e^{ - {\text 2}b{r_{m{l}}}}}({\text 1} - {e^{ - {\text 2}aD_{v,g}}})$ are derived in Appendix A and given by $\mathbb{E}[{D_{v,g}}]=\frac{T}{{M({\text 1} - \bar \varepsilon )}}$ and $ \mathbb{E}[{e^{ - {\text 2}b{r_{m{l}}}}}({\text 1} - {e^{ - {\text 2}aD_{v,g}}})]=\sum\limits_{n = {\text{1}}}^M {\frac{{{e^{ - {\text{2}}b{r_{mn}}}}{\Psi _n}}}{{M({\text{1 }} - {e^{ - \text2ah}}\bar \varepsilon )}}} $, respectively. Similar to the synchronous transmission case, an approximation for  $\overline {{\text{MSE}}} _m^{{\text{Asyn-infer}}} $ with respect to the MSSC is given by
\begin{align}
\overline {{\text{MSE}}} _{m,{\text{appr}}}^{{\text{Asyn-infer}}} = \overline {{\text{MSE}}} _m^{{\text{Asyn-infer}}}\left| {_{{e^{ - \text2b{r_{mn}}}} = \bar \rho _m^{{\text{spatial}}},n\ne m}}. \right.
\label{mse_asyn_appr}
\end{align}The tightness of (\ref{mse_asyn_appr}) is verified by Fig. \ref{fig_MSE_N} in the simulation results.

\subsection{Relationship Between the Average MSE of Reconstruction and Average BLEP}

From (\ref{mse_syn}) and (\ref{mse_asyn}), the relationship between the average MSE of reconstruction and average BLEP is analyzed in Proposition 2 and Lemma 2.

\textbf{Proposition 2.}  $\overline {\text{MSE}} _{m}^{{\text{Syn-infer}}}$ is mono-increasing with respect to average BLEP $\bar\varepsilon$. $\overline {\text{MSE}} _{m}^{{\text{Asyn-infer}}}$ is decreasing first and then increasing with respect to average BLEP $\bar\varepsilon$ if time shift $h\ne\frac{T}{M}$ and MSSC $\bar \rho _m^{{\text{spatial}}} < \Upsilon $, and mono-increasing otherwise, where $\Upsilon =  \frac{{(\text1 - {e^{ - \text2a(T-Mh)}})({e^{ - \text2b{r_{m(M - \text1)}}}} - {e^{ - \text2b{r_{mM}}}}(\text1 - {e^{ - \text2ah}}))}}{{{e^{\text2ah}}(M - \text1){{(\text1 - {e^{ - \text2ah}})}^{\text2}}}} - \frac{\text1}{{M - \text1}}$.

\emph{Proof}: Please refer to Appendix B.
$\hfill\blacksquare$

From Proposition 2, we have the following observations. a) A lower BLEP can reduce the average MSE of reconstruction under the inference with synchronous transmission scheme. This is due to that less packet loss occurs at the target sensor, leading to a reduced average MSE of reconstruction. b) A lower BLEP increases the average MSE of reconstruction under the inference with asynchronous transmission scheme when time shift $h\ne\frac{T}{M}$, MSSC $\bar \rho _m^{{\text{spatial}}} < \Upsilon $ and average BLEP $\bar\varepsilon\in(\text0,{{{\bar \varepsilon }^*}})$, where ${{{\bar \varepsilon }^*}}$ is the root of $\frac{{\partial \overline {{\text{MSE}}} _m^{{\text{Asyn-infer}}}}}{{\partial \bar \varepsilon }}=$ 0. This seems counterintuitive at first. The reason is that with a small time shift, the benefit of AoI reduction by asynchronous transmission is small. Meanwhile, with a lower BLEP, the frequency of inference increases in the asynchronous inference scheme. Accordingly, the inference error increases significantly with a relatively weak MSSC, thereby resulting in an increased average MSE of reconstruction.

\textbf{Lemma 2.} The upper and lower bounds of $\overline {\text{MSE}} _{m}^{{\text{Syn-infer}}}$ with respect to average BLEP $\bar\varepsilon$ are respectively given by  $\overline {\text{MSE}} _{m,{\text{ub},\upvarepsilon}}^{{\text{Syn-infer}}}=\sigma _{{\text{X,}}m}^{\text 2}$ and $\overline {\text{MSE}} _{m,{\text{lb},\upvarepsilon}}^{{\text{Syn-infer}}}= \sigma _{{\text{X,}}m}^{\text 2} -\frac{ \sigma _{{\text{X,}}m}^{\text 2}{{\gamma _{\text o}}{e^{ - {\text 2}a\tau }}({\text 1} - {e^{ - {\text 2}aT}})}}{{{\text 2}aT({\gamma _{\text o}} + {\text 1})}}$.
The upper and lower bounds of $\overline {\text{MSE}} _{m}^{{\text{Asyn-infer}}}$ with respect to average BLEP $\bar\varepsilon$ are respectively given by  $\overline {\text{MSE}} _{m,{\text{ub},\upvarepsilon}}^{{\text{Asyn-infer}}} =\sigma _{{\text{X,}}m}^{\text 2}$ and 
\begin{align}
\overline {{\text{MSE}}} _{m,{\text{lb},\upvarepsilon}}^{{\text{Asyn-infer}}} = \left\{ \begin{array}{l}
\overline {{\text{MSE}}} _m^{{\text{Asyn-infer}}}\left| {_{\bar \varepsilon  = {{\bar \varepsilon }^*}}} \right., h \ne \frac{T}{M}{\text{and }}\bar \rho _m^{{\text{spatial}}} < \Upsilon ,\\
\overline {{\text{MSE}}} _m^{{\text{Asyn-infer}}}\left| {_{\bar \varepsilon  =\text0}} \right.,\; \;    \text{otherwise},\\
\end{array} \right.\nonumber
\end{align}where $\overline {{\text{MSE}}} _m^{{\text{Asyn-infer}}}\left| {_{\bar \varepsilon  =\text0}} \right.=\sigma _{\text{X},m}^{\text2} - \frac{{\sigma _{\text{X},m}^{\text{2}}{\gamma _{\text{o}}}{e^{ - {\text2a}\tau }}{\alpha ^{{\text{Asyn}}}}}}{{{\text{2}aT}({\gamma _{\text{o}}}{+\text{ 1}})}}$, with
${\alpha ^{{\text{Asyn}}}} $\\$= {e^{ - \text2b{r_{mM}}}}{e^{ - \text2ah}}(\text1 - {e^{ - \text2a(T - Mh)}}) + (\text1 - {e^{ - \text2ah}})\sum\limits_{n = \text1}^M {{e^{ - \text2b{r_{mn}}}}} $.

Lemma 2 indicates that the upper bound of the average MSE of reconstruction with respect to average BLEP is the same under both schemes, and is related to the sample variance ${\sigma _{{\text{X}},m}^{\text{2}}}$ only. The lower bound of the average MSE of reconstruction by inference with synchronous transmission scheme depends on the sample variance ${\sigma _{{\text{X}},m}^{\text{2}}}$, delay $\tau$ and transmission period $T$, independent of the sensor number $M$ and spatial correlation factor $e^{-br_{mn,n\ne m}}$. The lower bound of the average MSE of reconstruction by inference with asynchronous transmission scheme is affected by all parameters in a complex manner.

\subsection{Relationship Between the Average MSE of Reconstruction and Spatial Correlation}

The spatial correlation factor ${{e^{ - b{r_{mn,n \ne m}}}}}$ and MSSC $\bar\rho_m^{\text{spatial}}$ represent the individual and global spatial correlations, respectively. Their relationship to the average MSE of reconstruction is given in Proposition 3.

\textbf{Proposition 3.} $\overline {{\text{MSE}}} _{m}^{{\text{Syn-infer}}}$ and $\overline {{\text{MSE}}} _{m}^{{\text{Asyn-infer}}}$ are both mono-decreasing with respect to spatial correlation factor ${{e^{ - b{r_{mn,n \ne m}}}}}$ and MSSC $\bar\rho_m^{\text{spatial}}$.

\emph{Proof}: The first derivatives of $\overline {\text{MSE}} _{m}^{{\text{Syn-infer}}}$ and $\overline {\text{MSE}} _{m}^{{\text{Asyn-infer}}}$ with respect to spatial correlation factor are given by $\frac{{\partial \overline {{\text{MSE}}} _m^{{\text{Syn-infer}}}}}{{\partial {e^{ - b{r_{m\tilde s}}}}}} = \frac{{-\sigma _{{\text{X,}}m}^{\text{2}}{\gamma _{\text{o}}}{e^{ - {\text{2}}a\tau }}({\text{1}} - {e^{ - {\text{2}}aT}})({\text{1}} - \bar \varepsilon ){{\bar \varepsilon }^{\tilde s - {\text{1}}}}{e^{ - b{r_{m\tilde s}}}}}}{{aT({\gamma _{\text{o}}} + {\text{1}})({\text{1}} - {e^{ - {\text{2}}aT}}{{\bar \varepsilon }^M})}}<$ 0 and $\frac{{\partial \overline {{\rm{MSE}}} _m^{{\text{Asyn-infer}}}}}{{\partial {e^{ - b{r_{mn,n\ne m}}}}}} =  \frac{{-\sigma _{{\text{X}},m}^{\text{2}}{\gamma _{\text{o}}}{e^{ - {\text{2}}a\tau }}({\text{1}} - \bar \varepsilon ){e^{ - b{r_{mn}}}}{\Psi _n}}}{{aT({\gamma _{\text{o}}} + {\text{1}})({\text{1}} - {e^{ - {\text{2}}ah}}\bar \varepsilon )}}< $ 0, respectively. Since the MSSC is a sum function of ${{e^{ - \text2b{r_{mn,n \ne m}}}}}$ and mono-increasing with respect to ${{e^{ - b{r_{mn,n \ne m}}}}}$, we have $\frac{{\partial \overline {{\text{MSE}}} _m^{{\text{Syn-infer}}}}}{{\partial {\bar\rho_m^{\text{spatial}}}}}<$ 0 and $\frac{{\partial \overline {{\text{MSE}}} _m^{{\text{Asyn-infer}}}}}{{\partial {\bar\rho_m^{\text{spatial}}}}}<$ 0. Hence, Proposition 3 is guaranteed.
$\hfill\blacksquare$

Proposition 3 suggests that stronger spatial correlations can reduce the average MSE of reconstruction of the inference with synchronous/asynchronous transmission scheme, due to the reduction of inference error. Substituting ${{e^{ - b{r_{mn,n \ne m}}}}}=$ 0 and 1 into (\ref{mse_syn}) and (\ref{mse_asyn}) yields Lemma 3.

\textbf{Lemma 3.} The upper and lower bounds of $\overline {\text{MSE}} _{m}^{{\text{Syn-infer}}}$ with respect to spatial correlation factor ${{e^{ - b{r_{mn,n \ne m}}}}}$ (MSSC $\bar\rho_m^{\text{spatial}}$) are respectively given by $\overline {{\text{MSE}}} _{m,{\text{ub},\uprho^{\text{spatial}}}}^{{\text{Syn-infer}}}=\sigma _{{\text{X,}}m}^{\text{2}} -{\beta ^{{\text{Syn}}}}$ and $\overline {{\text{MSE}}} _{m,{\text{lb},\uprho^{\text{spatial}}}}^{{\text{Syn-infer}}} = \sigma _{{\text{X,}}m}^{\text{2}} - {\beta ^{{\text{Syn}}}}\frac{{\text1 - {{\bar \varepsilon }^M}}}{{{\text{1}} - \bar \varepsilon }}$, where ${\beta ^{{\text{Syn}}}} = \frac{{\sigma _{{\text{X,}}m}^{\text{2}}{\gamma _{\text{o}}}{e^{ - {\text{2}}a\tau }}({\text{1}} - {e^{ - {\text{2}}aT}})({\text{1}} - \bar \varepsilon )}}{{{\text{2}}aT({\gamma _{\text{o}}} + {\text{1}})({\text{1}} - {e^{ - {\text{2}}aT}}{{\bar \varepsilon }^M})}}$. 
The upper and lower bounds of $\overline {{\text{MSE}}} _{m}^{{\text{Asyn-infer}}}$ with respect to spatial correlation factor ${{e^{ - b{r_{mn,n \ne m}}}}}$ (MSSC $\bar\rho_m^{\text{spatial}}$) are respectively given by $\overline {{\text{MSE}}} _{m,{\text{ub},\uprho^{\text{spatial}}}}^{{\text{Asyn-infer}}}=\sigma _{{\text{X}},m}^{\text{2}} -{\beta ^{{\text{Asyn}}}}{\Psi _m} $ and $\overline {{\text{MSE}}} _{m,{\text{lb},\uprho^{\text{spatial}}}}^{{\text{Asyn-infer}}}=\sigma _{{\text{X}},m}^{\text{2}} -{\beta ^{{\text{Asyn}}}}\sum\nolimits_{n = \text1}^M {{\Psi _n}} $, where ${\beta ^{{\text{Asyn}}}} = \frac{{\sigma _{{\text{X}},m}^{\text{2}}{\gamma _{\text{o}}}{e^{ - {\text{2}}a\tau }}{e^{ - {\text{2}}ah}}({\text{1}} - \bar \varepsilon )}}{{{\text{2}}aT({\gamma _{\text{o}}} + {\text{1}})({\text{1}} - {e^{ - {\text{2}}ah}}\bar \varepsilon )}}$.

Lemma 3 reveals that the upper and lower bounds of the average MSE of reconstruction with respect to the spatial correlation are dependent of the average BLEP, and affected by the parameters in a complex manner. Based on Propositions 2$\sim$3 and Lemmas 2$\sim$3, the relationship between the  average MSE of reconstruction and average BLEP and spatial correlation is summarized in Table I.

\begin{table*}[htbp]  
    \centering
    \caption{Relationship between the Average MSE of Reconstruction and Average BLEP and Spatial Correlation}
    			\vspace{-5pt}
    \begin{tabular}{|c |l| l|}\hline
        \makecell{Scheme}     &     \makecell{\;\;\;\;\;\;Average MSE of reconstruction vs.\\  \;\;\;\;\;\;Average BLEP}                  &       \makecell{ Average MSE of reconstruction vs.\\Spatial correlation factor (MSSC)}                                    
                                           \\\hline\hline
        \multirow{3}{*}{\vspace{-8pt}\makecell{Inference with \\ syn. transmission}}& \makecell{\;\;\;\;\;\;\;\;\;\;Mono-increasing} &  \makecell{\;\;\;\;\;\;\;\;\;Mono-decreasing} \\\cline{2-3} &  \makecell{\;\;\;\;\;\;\;\;\;\;\;\;Upper bound: $\sigma _{{\text{X,}}m}^{\text 2}$}&  \makecell{ \;\;\;\;\;Upper bound: $\sigma _{{\text{X,}}m}^{\text{2}} -{\beta ^{{\text{Syn}}}}$}\\\cline{2-3} &  \makecell        {\;\;\;\;Lower bound: $ \sigma _{{\text{X,}}m}^{\text 2} -\frac{ \sigma _{{\text{X,}}m}^{\text 2}{{\gamma _{\text o}}{e^{ - {\text 2}a\tau }}({\text 1} - {e^{ - {\text 2}aT}})}}{{{\text 2}aT({\gamma _{\text o}} + {\text 1})}}$ }  & \makecell{\;\;\;\;Lower bound: $\sigma _{{\text{X,}}m}^{\text{2}} - {\beta ^{{\text{Syn}}}}\frac{{\text1 - {{\bar \varepsilon }^M}}}{{{\text{1}} - \bar \varepsilon }}$}                     
         \\\hline\hline
        \multirow{3}{*}{\vspace{-4pt}\makecell{Inference with\\  asyn. transmission}} & \makecell{ Decreasing first and then increasing if $h\ne\frac{T}{M}$\\ and $\bar \rho _m^{{\text{spatial}}} <\Upsilon $, and mono-increasing otherwise }      & \makecell{\;\;\;\;\;\;\;\;\;Mono-decreasing}        \\\cline{2-3} &   \makecell{\;\;\;\;\;\;\;\;\;\;\;\;Upper bound: $\sigma _{{\text{X,}}m}^{\text 2}$}  & \makecell{\;\;\;Upper bound: $\sigma _{{\text{X}},m}^{\text{2}} -{\beta ^{{\text{Asyn}}}}{\Psi _m} $}     \\\cline{2-3} & \makecell{\;\;\;\;\;\;\;\;\;\;\;\;Lower bound: $\overline {{\text{MSE}}} _{m,{\text{lb},\upvarepsilon}}^{{\text{Asyn-infer}}}$ }  &  \makecell{Lower bound: $\sigma _{{\text{X}},m}^{\text{2}} -{\beta ^{{\text{Asyn}}}}\sum\nolimits_{n = \text1}^M {{\Psi _n}} $}                                                          \\\hline
    \end{tabular}
    \label{simplt}
\end{table*}

	\begin{table*}[t]    
			\centering
			\caption{Preference Regions Comparison}
			\vspace{-5pt}
			\renewcommand\arraystretch{1}{
				\begin{tabular}{|c|c|c|}
					\hline
					&\makecell{Comparison of average MSE of reconstruction}&\makecell{Condition}\\
					\hline\hline
					\makecell {\vspace{-2.5pt}Case 1}& \makecell{No Inference superior\\($\overline {{\text{MSE}}} _{m}^{{\text{No infer}}}\le\overline {{\text{MSE}}} _m^{{\text{Syn-infer}}}< \overline {{\text{MSE}}} _m^{{\text{Asyn-infer}}}$)}& \makecell{$\bar \rho _{m}^{{\text{spatial}}}\le\bar \rho _{m,\text{thr1}}^{{\text{spatial}}}$}\\
					\cline{2-3}
					\hline\hline
					\makecell{\vspace{-2.5pt}Case 2}& \makecell{Inference with syn. transmission superior\\($\overline {{\text{MSE}}} _{m}^{{\text{Syn-infer}}}<\overline {{\text{MSE}}} _m^{{\text{No infer}}} \;\text{and}\; \overline {{\text{MSE}}} _m^{{\text{Asyn-infer}}}$)}&\makecell{$\bar \rho _{m,\text{thr1}}^{{\text{spatial}}}<\bar \rho _{m}^{{\text{spatial}}}<\bar \rho _{m,\text{thr2}}^{{\text{spatial}}}$}
					\\
					\hline
										\cline{2-3}
					\hline\hline
					\makecell{\vspace{-2.5pt}Case 3}& \makecell{Inference with asyn. transmission  superior\\($\overline {{\text{MSE}}} _{m}^{{\text{Asyn-infer}}}\le\overline {{\text{MSE}}} _m^{{\text{Syn-infer}}}< \overline {{\text{MSE}}} _m^{{\text{No infer}}}$)}&\makecell{$\bar \rho _{m}^{{\text{spatial}}}\ge\bar \rho _{m,\text{thr2}}^{{\text{spatial}}}$}
					\\
					\hline
			\end{tabular}}
		\end{table*}

\section{Preference Regions Analysis}

The inference with synchronous transmission scheme enables a high probability of successful reception at the cloud server, and meanwhile, the inference with asynchronous transmission scheme enables a low AoI by providing consecutive fresh samples. Both schemes can achieve a superior average MSE performance of reconstruction. Hence, we analyze the preference regions of the two schemes in terms of the MSSC.

Due to the complex form of the closed-form expressions $\overline {{\text{MSE}}} _m^{{\text{Syn-infer}}}$ and $\overline {{\text{MSE}}} _m^{{\text{Asyn-infer}}}$, we use their tight approximations $\overline {{\text{MSE}}} _{m,\text{appr}}^{{\text{Syn-infer}}}$ and $\overline {{\text{MSE}}} _{m,\text{appr}}^{{\text{Asyn-infer}}}$ as the analysis objects. Compared to the inference with synchronous transmission scheme, the inference with asynchronous transmission scheme achieves the performance gain of average MSE of reconstruction as $ \frac{{\overline {{\text{MSE}}} _m^{{\text{Syn-infer}}}}}{{\overline {{\text{MSE}}} _m^{{\text{Asyn-infer}}}}} \approx \frac{{\overline {{\text{MSE}}} _{m,{\text{appr}}}^{{\text{Syn-infer}}}}}{{\overline {{\text{MSE}}} _{m,{\text{appr}}}^{{\text{Asyn-infer}}}}} $, which is larger than one with MSSC $\bar \rho _m^{{\text{spatial}}}>\bar \rho _{m,{\text{thr2}}}^{{\text{spatial}}}$, where $\bar \rho _{m,{\text{thr2}}}^{{\text{spatial}}} = \frac{{\mathchar'26\mkern-10mu\lambda  - {\Psi _m}}}{{\sum\limits_{n = {\text{1},}n \ne m}^M {{\Psi _n}}  - \mathchar'26\mkern-10mu\lambda (\bar \varepsilon  - {{\bar \varepsilon }^M})/({\text1 -}\bar \varepsilon )}}$ and $\mathchar'26\mkern-10mu\lambda  = \frac{{{\text{(1}} - {e^{ - {\text{2}}aT}}){\text{(1}} - {e^{ - {\text{2}}ah}}\bar \varepsilon )}}{{{\text{1}} - {e^{ - {\text{2}}aT}}{{\bar \varepsilon }^M}}}$.
This yields Proposition 4.

\textbf{Proposition 4.} The inference with asynchronous transmission scheme can achieve a lower average MSE of reconstruction than the inference with synchronous transmission scheme as long as the MSSC $\bar \rho _m^{{\text{spatial}}} >\bar \rho _{m,{\text{thr2}}}^{{\text{spatial}}}$.

Based on Propositions 4 and 1, the performance comparison of average MSE of reconstruction among the no inference, inference with synchronous transmission and inference with asynchronous transmission schemes is summarized in Table II, and their preference regions regarding the MSSC are concluded as well. They provide a clear guidance on the spatial correlation-aware transmission mode selection between synchronous and asynchronous, and reconstruction method selection between inference and non-inference. Generally, the inference with synchronous transmission scheme is preferable with a relatively weak MSSC. The inference with asynchronous transmission scheme is desirable in a relatively strong MSSC regime.

Note that although the typical separable spatial-temporal correlation model with exponential form in (1) is considered in our analysis. Our results can be extended to other non-separable and non-linear spatial-temporal correlation models, by following a similar analytical method in this paper. A more detailed discussion for the MSEs with different temporal-spatial correlation models is given in \cite{Hribar}, \cite{Hristopulos} and \cite{Porcu}.

\section{Adaptations of Blocklength and Time Shift}

For the inference with synchronous/asynchronous transmission scheme, a short blocklength can reduce the MSE of reconstruction by reducing transmission delay and AoI. However, a too short blocklength increases BLEP \cite{Yu}, which impairs the MSE performance of reconstruction. Also, the length of time shift in the asynchronous transmission affects the tradeoff between the intra-period AoI and the inter-period AoI. 
Hence, based on the closed-form average MSE expressions derived in Section III, we investigate blocklength and time shift adaptations to minimize the average MSE of reconstruction for the inference with synchronous transmission and inference with asynchronous transmission schemes.

\subsection{Blocklength Adaptation for Inference with Synchronous Transmission Scheme}
Based on (\ref{mse_syn}), the optimization problem for the inference with synchronous transmission scheme is formulated as
\begin{align}
\textbf{P1}: \;\;& \mathop {\min }\limits_{N} \;\;  \overline {{\text{MSE}}} _m^{\text{Syn-infer}}\\
&\;   \text{s.t.}\;\; \; (\text{C1}): N_{\text{min}}\le N\le T/T_{\text s},  N \in {{\mathbb{ N}}_ + },\nonumber 
\end{align}where (C1) constraints the minimum and maximum allowable blocklengths. Relaxing the integer constraint of $N$ into continuous space, we provide Proposition 5 and Lemma 4.

\textbf{Proposition 5.} Given information bits $L\ge\pi$, optimization problem \textbf{P1} is convex with respect to blocklength $N$.

\emph{Proof}: Please refer to Appendix C.
$\hfill\blacksquare$

\textbf{Lemma 4.} Given information bits $L\ge\pi$, the optimal blocklength $N_*^{\text{Syn-infer}}$ for Problem \textbf{P1} is given by
\begin{equation}
\begin{aligned}
N_{*}^{{\text{Syn-infer}}} = \left\{ \begin{array}{l}
N_{\text{min} },\;\;\;\;\;\;\;\;\;\;\;\;\;\;\;\;\;\;\;\;\;\;\;\;\;\;  H(N_{\text{min} }) > {\text 0},\\
N_{\text{max} }, \;\;\;\;\;\;\;\;\;\;\;\;\;\;\;\;\;\;\;\;\;\;\;\;\;  H(T/T_{\text s}) < {\text 0},\\
\mathop {\arg \min }\limits_{N\in \left\{ {\left\lfloor {\hat N} \right\rfloor ,\left\lceil {\hat N} \right\rceil } \right\}} \overline {{\text{MSE}}} _m^{{\text{Syn-infer}}},\;{\text{otherwise}},
\end{array} \right.
\end{aligned}
\end{equation}where ${\hat N}$ is the root of $H(N)={\text 0}$, with $H(N) \buildrel \Delta \over = \frac{{\partial \overline {{\text{MSE}}} _m^{{\text{Syn-infer}}}}}{{\partial N}} = \frac{{\sigma _{{\text{X}},m}^{\text 2}{\gamma _{\text{o}}}{e^{ - {\text 2}a\tau }}({\text 1} - {e^{ - {\text 2}aT}})}}{{{\text 2}aT({\gamma _{\text{o}}} + {\text 1}){{({\text 1} - {e^{ - {\text 2}aT}}{{\bar \varepsilon }^M})}^{\text 2}}}} \sum\limits_{\tilde s = {\text{1}}}^M {{e^{ - {\text{2}}b{r_{m\tilde s}}}}{{\bar \varepsilon }^{\tilde s - {\text{2}}}}} [({\text{1}} - {e^{ - {\text{2}}aT}}{{\bar \varepsilon }^M})(\frac{{\partial \bar \varepsilon }}{{\partial N}} + \text2a{T_{\text{s}}}\bar \varepsilon (\text1 - \bar \varepsilon )) - \frac{{\partial \bar \varepsilon }}{{\partial N}}({\text{1}} - \bar \varepsilon )({e^{ -\text 2aT}}{{\bar \varepsilon }^M}(M - \tilde s) + \tilde s)]$ and $\frac{{\partial {{\bar \varepsilon }}}}{{\partial N}} \approx \frac{{(\sqrt {\pi L}  - L{e^{L/N}}){e^{ - \frac{{\text 1}}{{{{\bar \gamma }_{{\text r}}}}}(\eta  - \sqrt {\pi L} /N)}}}}{{{{\bar \gamma }_{{\text r}}}{N^{\text 2}}}}$ \cite{Yu}.

\begin{algorithm}[h!]   
\caption{The JTSBO Algorithm }\label{[1]}
\text{1}\;\,{\bf Initialize:}\;Iteration index $i=$ 1, maximum number of iterations $I_{\text {max}}$, time shift $h^{(i)}$, blocklength $N^{(i)}$, and error tolerances $\widetilde \theta_{\text h}$ and $\widetilde \theta_{\text N}$;\\
\text{2}\;\,\textbf{Repeat}\\
\text{3}\;\,Calculate $h^{(i)}$ using (\ref{h}) with a given $N^{(i-\text1)}$, and then calculate $N^{(i)}$ using (\ref{N}) with given $h^{(i)}$;\\
\text{4}\;\,$i=i+\text1$;\\
\text{5}\;\,\textbf{Until} $i\ge I_{\text {max}}$ or $|h^{(i)}-h^{(i-\text1)}|<\widetilde \theta_{\text h}$ $\&$ $|N^{(i)}-N^{(i-\text1)}|<\widetilde \theta_{\text N}$\\
\text{6}\;\,\textbf{Output:}\;  $h_*^{\text{Asyn-infer}}$ and $N_*^{\text{Asyn-infer}}$. 
\end{algorithm}

\subsection{Time Shift and Blocklength Adaptations for Inference with Asynchronous Transmission Scheme}
Based on (\ref{mse_asyn}), the optimization problem for the inference with asynchronous transmission scheme is formulated as 
\begin{align}
\textbf{P2}: \;\;& \mathop {\min }\limits_{N,h} \;\;  \overline {{\text{MSE}}} _m^{\text{Asyn-infer}}\\
&\;   \text{s.t.}\;\;  (\text{C2}): N_{\text{min}}T_{\text s}\le NT_{\text s}\le T-(M-\text1)h, N \in {{\mathbb{ N}}_ + },\nonumber  \\
&\;\;\;\;\;\;\;\;\;\;\;\;\;\;\;\;\; T_{\text s}\le h\le \frac{{T - \tau }}{{M - \text1}}.\nonumber 
\end{align}Problem \textbf{P2} is not jointly convex with respect to both time shift and blocklength. To address this issue, we decompose it as
\begin{align}
\textbf{P2.1}: \;\; \mathop {\min }\limits_{h} \;\;  \overline {{\text{MSE}}} _m^{\text{Asyn-infer}}\;\;\;  \text{s.t.}\;\;  (\text{C2}), \\
\textbf{P2.2}: \;\; \mathop {\min }\limits_{N} \;\;  \overline {{\text{MSE}}} _m^{\text{Asyn-infer}}\;\;\;  \text{s.t.}\;\;  (\text{C2}),
\end{align}and provide Proposition 6.

\textbf{Proposition 6.} Given blocklength $N$, Problem \textbf{P2.1} is convex with respect to time shift $h$. Given information bits $L\ge \pi$ and time shift $h=T/M$, Problem \textbf{P2.2} is convex with respect to blocklength $N$.

\emph{Proof}: Please refer to Appendix D.
$\hfill\blacksquare$

In the scenario with time shift $h\ne T/M$, the convexity of $\overline {{\text{MSE}}} _m^{{\text{Asyn-infer}}}$ with respect to blocklength $N$ is verified in Fig. \ref{fig_MSE_N}. Based on Proposition 6, we can provide Lemma 5.

\textbf{Lemma 5.} Given blocklength $N$, the optimal time shift $h_*^{\text{Asyn-infer}}$ for Problem \textbf{P2.1} is given by
\begin{equation}
\begin{aligned}
h_{*}^{{\text{Asyn-infer}}} = \left\{ \begin{array}{l}
T_{\text s},\;\;\;\;\;\;\;\;\;\;\;\;\;\;\;\;\;\;\;\;\;\;\;\;\;\;\;\;\;\;\;\;\;\, J(T_{\text s}) > {\text 0},\\
\frac{{T - \tau }}{{M - \text1}}, \;\;\;\;\;\;\;\;\;\;\;\;\;\;\;\;\;\;\;\;\;\;\;\;\;\,     J(\frac{{T - \tau }}{{M - \text1}}) < {\text 0},\\
\mathop {\arg \min }\limits_{h\in \left\{ {\left\lfloor {\hat h} \right\rfloor ,\left\lceil {\hat h} \right\rceil } \right\}} \overline {{\text{MSE}}} _m^{{\text{Asyn-infer}}},\;{\text{otherwise}}.
\end{array} \right.
\label{h}
\end{aligned}
\end{equation}Given time shift $h$ and information bits $L\ge \pi$, the optimal blocklength $N_*^{\text{Asyn-infer}}$ for Problem \textbf{P2.2} is given by
\begin{equation}
\begin{aligned}
N_{*}^{{\text{Asyn-infer}}} = \left\{ \begin{array}{l}
N_{\text{min} },\;\;\;\;\;\;\;\;\;\;\;\;\;\;\;\;\;\;\;\;\;\;\;\;\;\;\;  F(N_{\text{min} }) > {\text 0},\\
\frac{{T - (M - \text1)h}}{{{T_{\text{s}}}}}, \;\;\;\;\;\;\;\;\;\;  F(\frac{{T - (M - \text1)h}}{{{T_{\text{s}}}}}) < {\text 0},\\
\mathop {\arg \min }\limits_{N\in \left\{ {\left\lfloor {\bar N} \right\rfloor ,\left\lceil {\bar N} \right\rceil } \right\}} \overline {{\text{MSE}}} _m^{{\text{Asyn-infer}}},{\text{otherwise}}.
\end{array} \right.
\label{N}
\end{aligned}
\end{equation}The ${\hat h}$ is the root of $J(h)={\text 0}$, with $J(h) \buildrel \Delta \over = \frac{{\partial \overline {{\text{MSE}}} _m^{{\text{Asyn-infer}}}}}{{\partial h}}= - \frac{{\sigma _{{\text{X}},m}^{\text{2}}{\gamma _{\text{o}}}{e^{ - {\text{2}}a\tau }}{{({\text{1}} - \bar \varepsilon )}^{\text2}}{e^{ - {\text{2}}ah}}}}{{T({\gamma _{\text{o}}} + {\text{1}}){{({\text{1}} - {e^{ - {\text{2}}ah}}\bar \varepsilon )}^{\text2}}}}\sum\limits_{n = {\text{1}}}^M {{e^{ - {\text{2}}b{r_{mn}}}}}     \left( {\text1 - \frac{{{{\bar \varepsilon }^{M - n}}{e^{\text2ahn}}}}{{{\text{1}} - {e^{ - {\text{2}}aT}}{{\bar \varepsilon }^M}}}} \right.    \left( ( \right.\text1 - {e^{ - {\text{2}}ah}}\bar \varepsilon )M{e^{ - {\text{2}}ahM}} + (\text1 - n(\text1 - {e^{ - {\text{2}}ah}}\bar \varepsilon ))({e^{ - {\text{2}}ahM}} - {e^{ - {\text{2}}aT}})))$. 
The $\bar N$ is the root of $F(N)=$ 0, with $F(N) \buildrel \Delta \over = \frac{{\partial \overline {{\text{MSE}}} _m^{{\text{Asyn-infer}}}}}{{\partial N}}=$\\$- \frac{{\sigma _{{\text{X}},m}^{\text{2}}{\gamma _{\text{o}}}{e^{ - {\text{2}}a\tau }}}}{{\text2aT({\gamma _{\text{o}}} + {\text{1}}){{({\text{1}} - {e^{ - {\text{2}}ah}}\bar \varepsilon )}^{\text2}}}}   \sum\limits_{n = {\text{1}}}^M {{e^{ - {\text{2}}b{r_{mn}}}}}      [ ({\text{1}} - \bar \varepsilon )({\text{1}} - {e^{ - {\text{2}}ah}}\bar \varepsilon )(\frac{{\partial \Psi_n}}{{\partial N}} $ $- {\text{2}}a{T_{\text{s}}}\Psi_n) - \frac{{\partial \bar \varepsilon }}{{\partial N}}\Psi_n(\text1 - {e^{ - {\text{2}}ah}})]$ and $\frac{{\partial \Psi_n}}{{\partial N}} = \frac{{({e^{ - {\text{2}}ahM}} - {e^{ - {\text{2}}aT}}){{\bar \varepsilon }^{M - n - \text1}}}}{{{e^{ - \text2ah(n+\text1)}}{{{\text{(1}} - {e^{ - {\text{2}}aT}}{{\bar \varepsilon }^M})}^{\text2}}}}\frac{{\partial \bar \varepsilon }}{{\partial N}}(M({\text{1}} - \bar \varepsilon ) - (n({\text{1}} - \bar \varepsilon ) + \bar \varepsilon ){\text{(1}} - {e^{ - {\text{2}}aT}}{{\bar \varepsilon }^M}))$.

With Lemma 5, Problem \textbf{P2} can be solved in an iterative manner, as described in the proposed JTSBO algorithm in Algorithm 1. The complexity of the proposed JTSBO algorithm depends on the maximum number of iterations $I_{\text{max}}$ and the complexity of solving the non-linear equations of $J(h) =$ 0 and $F(N) =$ 0. Using Newton’s method \cite{YuB} to solve the two equations, the total complexity is $O({I_{\max }}(||{u_{\text{h}}}|| + ||{u_{\text{N}}}||))$, where ${u_{\text{h}}}$ and ${u_{\text{N}}}$ are the gaps between the initialized solution and the corresponding exact solution of the two equations, respectively. In contrast, exhaustive search for the inference with asynchronous transmission scheme requires a much higher complexity of $O\left( {\frac{{({N_{\text{max} }} - {N_{\text{min} }})(\text2T/{T_{\text{s}}} - {N_{\text{max} }} - {N_{\text{min} }})}}{{\text2(M -\text1)}}} \right)$, with ${N_{\text{max}}} =  - (M - \text1) + T/{T_{\text{s}}}$.

\begin{figure}[!]    
\centering
\subfigure[]{
\includegraphics[height=3.065cm]{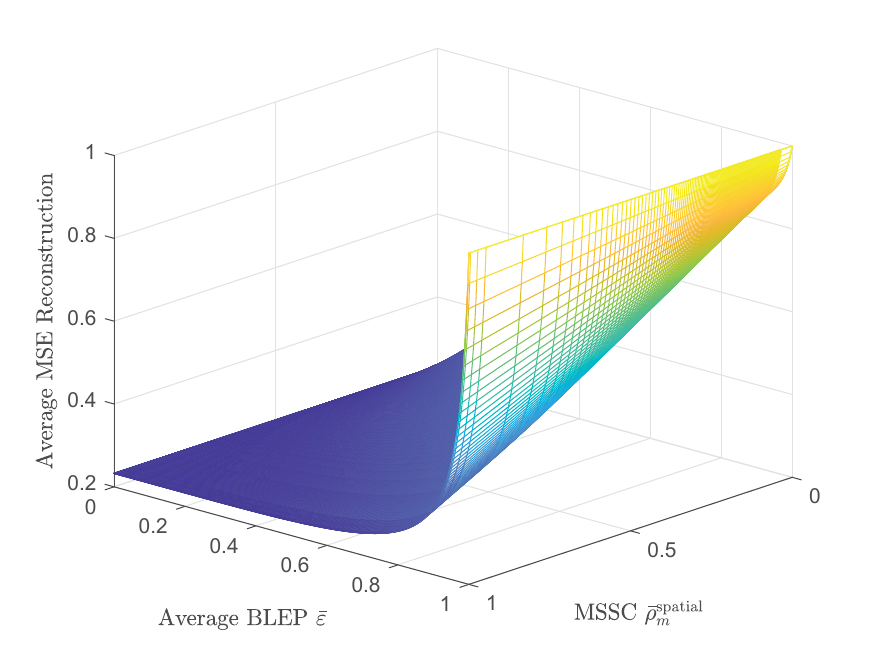} 
}
\hspace{-15pt}  
\subfigure[]{
\includegraphics[height=3.065cm]{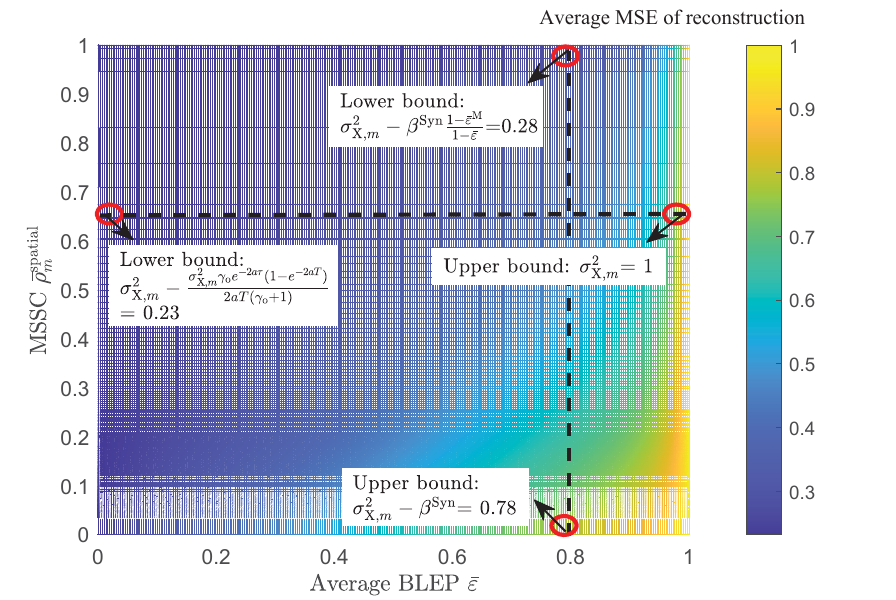}
}
\caption{Joint impact of average BLEP and MSSC on average MSE of reconstruction by inference with synchronous transmission scheme at transmission period $T=150$ ms.} 
\label{3d_syn} 
\end{figure}

\begin{figure}[!]    
\centering
\subfigure[]{
\includegraphics[height=3.065cm]{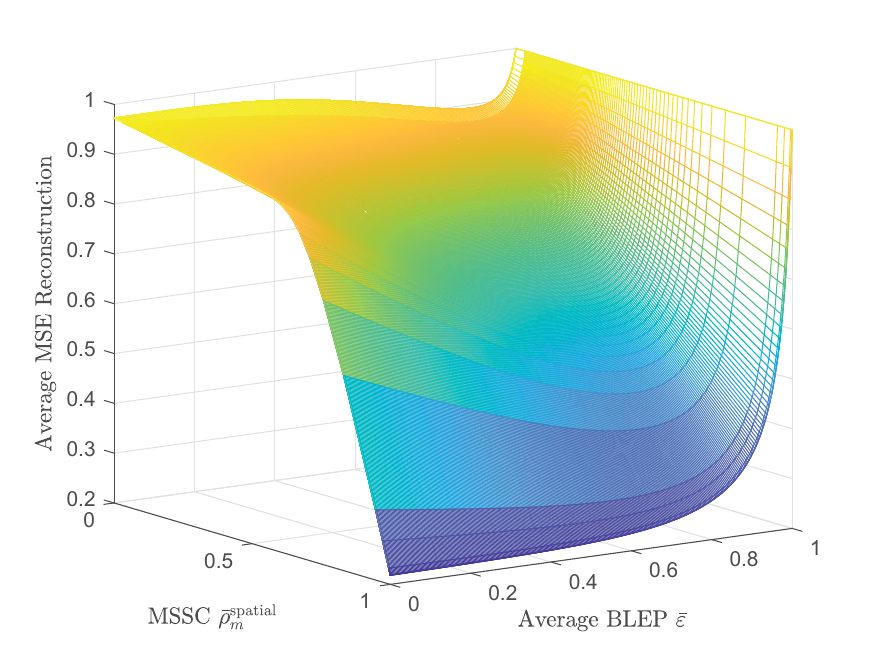}  
}
\hspace{-15pt}  
\subfigure[]{
\includegraphics[height=3.065cm]{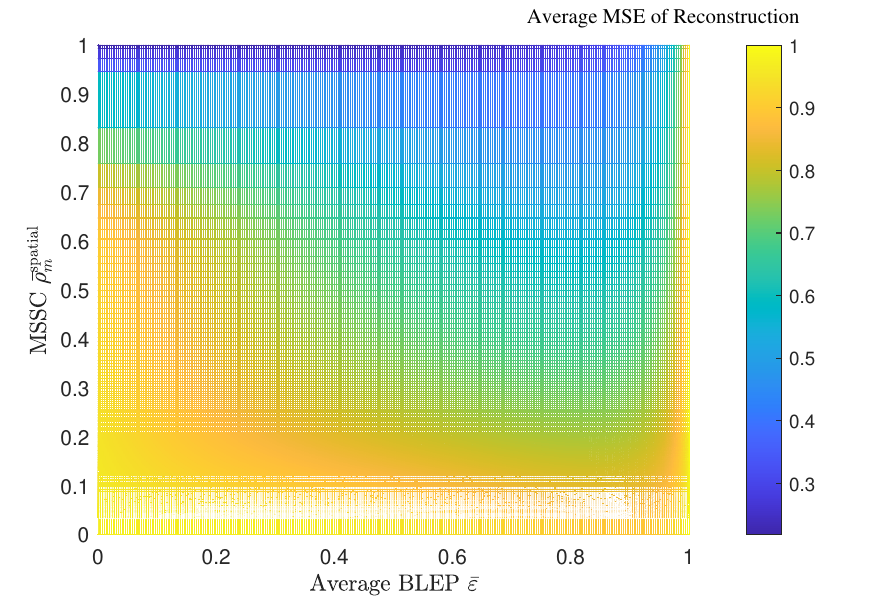}
}
\quad
\subfigure[]{
\includegraphics[height=3.065cm]{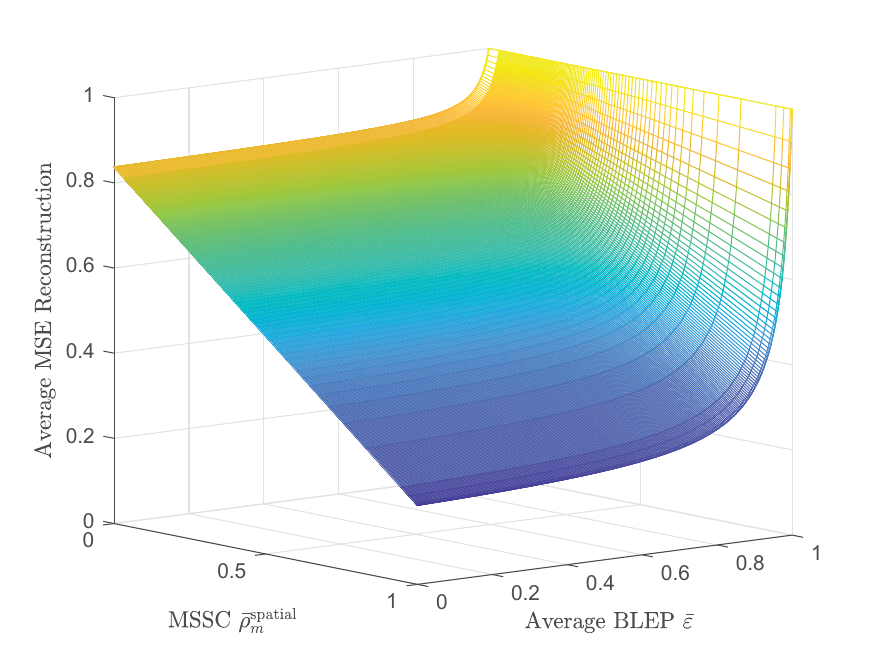}
}
\hspace{-15pt}  
\subfigure[]{
\includegraphics[height=3.065cm]{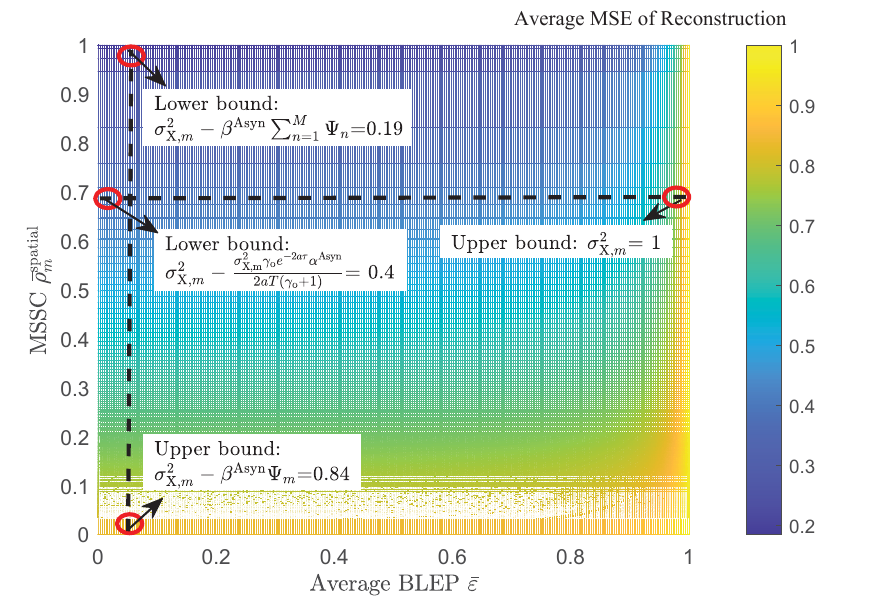}
}
\caption{Joint impact of average BLEP and MSSC on average MSE of reconstruction by inference with asynchronous transmission scheme at transmission period $T=150$ ms and time shift (a)-(b) $h=$ 5 ms and (c)-(d) $h=\frac{T}{M}=$ 30 ms.} 
\label{3d_asyn} 
\end{figure}

\section{Simulation Results}

In this section, simulation results are presented to validate the analysis of average MSE of reconstruction, and evaluate the performance of the proposed blocklength and time shift adaptation methods. We consider $M=$ 5 sensors randomly distributed within a (20$\times$20) ${\text m}^{\text2}$ square area according to an HPPP with density $\lambda_{\text d}=M/(\pi R^{\text2})$, where $R=$ 10 m is the radius of the largest circle of the square area. The sample and noise variances are respectively set to $\sigma^{\text2}_{\text{X},m}=$ 1 and $\sigma^{\text 2}_{{\text V},m}=$ 0.2, yielding the observation SNR $\gamma_{\text o}=\frac{{\sigma _{{\text{X},m}}^{\text 2}}}{{\sigma _{{\text{V},m}}^{\text 2}}}=$ 5. The scaling factors are set to $b=$ 0.01, except for Figs. 4$\sim$6, and $a=$ 0.5 for Figs. 4 and 5, and $a=$ 2 for Figs. 6$\sim$12. We set the information bits per noisy sample $L=$ 160 bits and the blocklength $N=$ 80 c.u.. The bandwidth per sensor is 10 KHz. Hence, the symbol duration $T_{\text s}=$ 0.1 ms and transmission delay $\tau=NT_{\text s}=$ 8 ms. The maximum number of iterations in the proposed JTSBO algorithm is set to 3. The simulation results are obtained by Monte Carlo simulation with $\text{10}^{\text5}$ realizations of periodic transmission. The distance between the cloud server and sensors is assumed to be the same, as $d =$ 200 m. The pathloss is modeled as 35.3 + 37.6$\text{log}_{\text{10}}(d) + W$, where $W$ is the shadowing which follows log normal distribution with zero mean and standard deviation of 8 \cite{Hou}. The received noise power spectrum density is $-$174 dBm/Hz \cite{Hou}. Transmission power at each sensor is $P_{\text t}=$ 0.2 mW, yielding the average received SNR $\bar{\gamma _{\text{r}}} = \frac{{\beta {P_{\rm{t}}}}}{{\sigma _{\text{r}}^{\text{2}}}}=$ 5 dB. For brevity, the inference with synchronous transmission and the inference with asynchronous transmission schemes are labeled as `Syn-infer' and `Asyn-infer', respectively. We focus on the performance of a certain sensor, and the representation of `sensor $\text{S}_m$' is omitted.

\vspace{-5pt}
\subsection{Average MSE Analysis}

Figs. \ref{3d_syn} and  \ref{3d_asyn} respectively show the joint impact of average BLEP and MSSC on average MSE of reconstruction by inference with synchronous transmission scheme and inference with asynchronous transmission scheme, where $b\in(\infty ,0)$ yielding MSSC $\bar \rho _m^{{\text{spatial}}}\in$ (0, 1). It is observed in Fig. \ref{3d_syn} that the average MSE of reconstruction by inference with synchronous transmission scheme is mono-increasing with respect to average BLEP. The average MSE of reconstruction by inference with asynchronous transmission scheme is decreasing first and then increasing with respect to average BLEP at time shift $h=$ 5 ms, and mono-increasing at time shift $h=\frac{T}{M}=$ 30 ms, as shown in Fig. \ref{3d_asyn}(a)-(b) and Fig. \ref{3d_asyn}(c)-(d), respectively. For both schemes, the average MSE of reconstruction is mono-decreasing  with respect to MSSC, and upper and lower bounded with respect to average BLEP and MSSC. These results validate the conclusions in Propositions 2$\sim$3 and Lemmas 2$\sim$3.

	\begin{figure}[t!]
	\centering
	{\includegraphics[height=5.5cm]{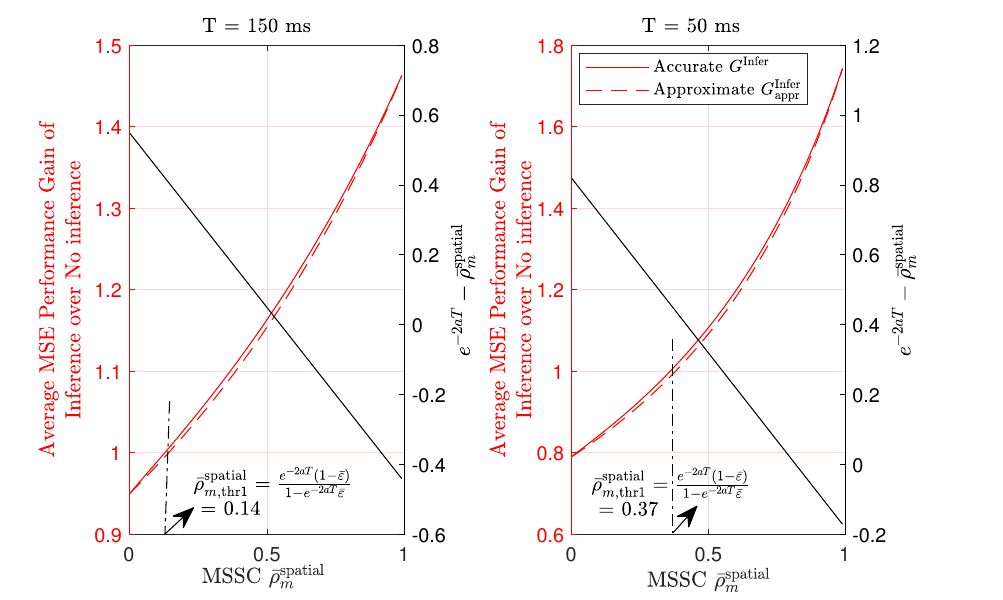}}
	\caption{Impact of the MSSC on the average MSE performance gain of inference over no inference and the gap of ${e^{ - \text2aT}} -\bar \rho _m^{{\text{spatial}}}$.}
	\label{fig_Gain_byinference_rou}
\end{figure}

Fig. \ref{fig_Gain_byinference_rou} shows the impact of the MSSC on the average MSE performance gain of inference over no inference and the gap of ${e^{ - \text2aT}} - \bar\rho _m^{{\text{spatial}}}$. The approximate gain $G^{\text{Infer}}_{\text{appr}}$ is tight to the accurate gain $G^{\text{Infer}}$. The inference achieves a significant performance gain of average MSE of reconstruction over the no inference case, even under weak MSSC and under weaker than the squared temporal correlation at a transmission period length, \emph{i.e.}, $\bar \rho _m^{{\text{spatial}}}<e^{ - {\text{2}}aT}$, as analyzed in Lemma 1. At transmission period $T=$ 150 and 50 ms, the MSSC threshold $\bar\gamma_{m,\text{thr1}}^{\text{spatial}}$ is 0.14 and 0.37, respectively. The inference achieves a performance gain with $\bar\gamma_m^{\text{spatial}}<e^{-\text2aT}$ at $T=$ 150 ms with $\bar \rho _m^{{\text{spatial}}}\in$ (0.14, 0.5) and $T=$ 50 ms with $\bar \rho _m^{{\text{spatial}}}\in$ (0.37, 0.8).

	\begin{figure}[t]  
	\centering
	{\includegraphics[height=5.5cm]{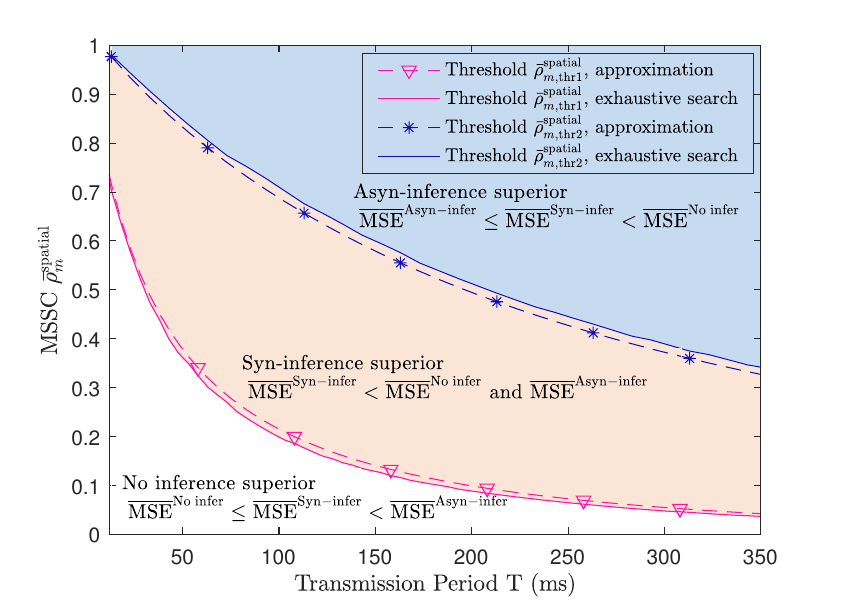}}
	\caption{Preference regions of the inference with synchronous transmission and inference with asynchronous transmission schemes in terms of the MSSC at time shift $h=\frac{{T - \tau }}{{M - \text1}}$ ms.}
	\label{fig_RoP_Thr_rou_T_three}
\end{figure}

	\begin{figure}[t!]  
	\centering
	{\includegraphics[height=5.5cm]{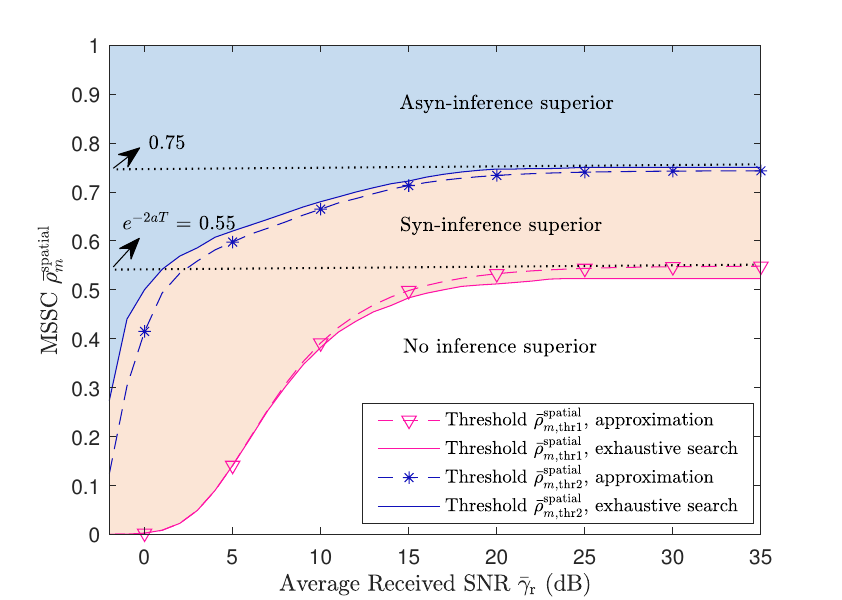}}
	\caption{Preference regions of the inference with synchronous transmission and inference with asynchronous transmission schemes in terms of the MSSC at transmission period $T=$ 150 ms and time shift $h=\frac{{T - \tau }}{{M - \text1}}$ ms.}
	\label{fig_RoP_Thr_rou_SNR_three}
\end{figure}

Fig. \ref{fig_RoP_Thr_rou_T_three} shows the preference regions of the inference with synchronous transmission and inference with asynchronous transmission schemes in terms of the MSSC under different transmission periods, and demonstrates the tightness of the approximate MSSC thresholds derived in Propositions 1 and 4. The approximate MSSC thresholds are close to the precise thresholds by exhaustive search, with a much lower complexity. The threshold $\bar\gamma^{\text{spatial}}_{{m,\text{thr2}}}$ is significantly larger than the threshold $\bar\gamma^{\text{spatial}}_{{m,\text{thr1}}}$. The reason is that the inference with synchronous transmission scheme always uses the latest data that has the strongest spatial correlation to the target sensor to infer, and hence allows a relatively weak MSSC to enable an average MSE performance gain. In contrast, the inference with asynchronous transmission scheme uses the data of every sensor equally, and therefore requires a relatively strong MSSC for achieving an average MSE performance gain. The MSSC thresholds decrease with a longer transmission period, as the benefit of AoI reduction by inference increases, leading to an increased tolerance in spatial correlation. In general, the inference with synchronous transmission scheme is preferable with relatively weak MSSC and medium transmission period.  The inference with asynchronous transmission scheme is desirable in a relatively strong MSSC and long transmission period regime.


Fig. \ref{fig_RoP_Thr_rou_SNR_three} shows the preference regions under different average received SNRs. The MSSC thresholds decrease with a lower average received SNR. The reason is that a lower average received SNR results in a lower BLEP, which leads to a more significant benefit of AoI reduction by inference. In addition, the MSSC thresholds saturate at high average received SNR. The inference with synchronous transmission scheme is desirable with a low-medium average received SNR. The inference with asynchronous transmission scheme is preferable with a low average received SNR.


	\begin{figure}[t!]
	\centering
	{\includegraphics[height=5.5cm]{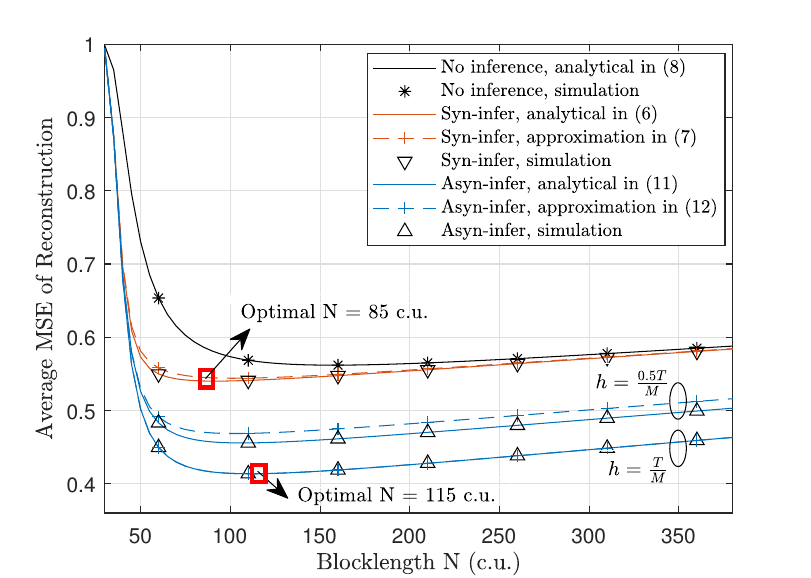}}
	\caption{Impact of blocklength on average MSE of reconstruction at transmission period $T=$ 300 ms and average received SNR $\bar\gamma_{\text{r}}=$ 15 dB.}\label{fig_MSE_N}
\end{figure}

	\begin{figure}[t]
	\centering
	{\includegraphics[height=5.5cm]{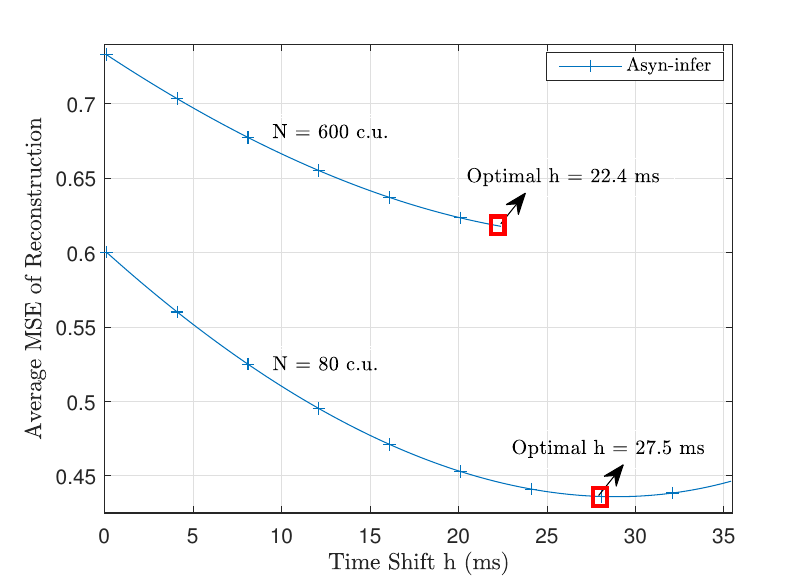}}
	\caption{Impact of time shift on average MSE of reconstruction by the inference with asynchronous transmission scheme at transmission period $T=$ 150 ms, average received SNR $\bar\gamma_{\text r}=$ 15 dB and $a=$ 4.}
	\label{fig_MSE_time_shift}
\end{figure}

Fig. \ref{fig_MSE_N} shows the impact of blocklength on average MSE of reconstruction and validates the analytical results. The analytical results of average MSE of reconstruction in (6), (8) and (11) match the simulation results, and the approximations in (7) and (12) are tight to the simulation results. There are optimal blocklengths to minimize the average MSE of reconstruction of the inference with synchronous transmission and  inference with asynchronous transmission schemes, respectively. Fig. \ref{fig_MSE_time_shift} shows the impact of time shift on average MSE of reconstruction by the inference with asynchronous transmission scheme. Given a blocklength, there is an optimal time shift to minimize the value of the average MSE of reconstruction. These results validate Propositions 5 and 6.

\subsection{Blocklength and Time Shift Adaptations}

	\begin{figure}[t!]  
	\centering  
	{\includegraphics[height=5.5cm]{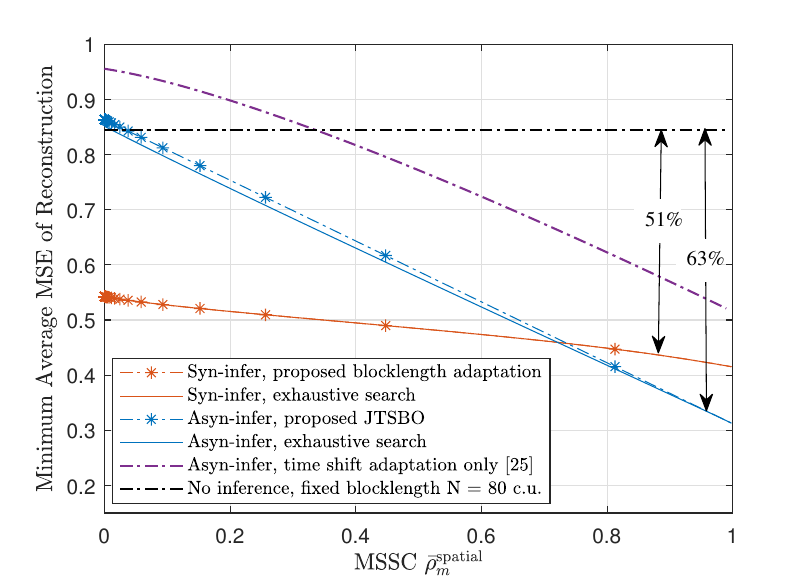}}
	\caption{Impact of the MSSC on the minimum average MSE of reconstruction by the proposed adaptation methods at transmission period $T=$ 150 ms and average received SNR $\bar\gamma_{\text r}=$ 5 dB}
	\label{fig_Min_MSE_space_corr}
\end{figure}

Fig. \ref{fig_Min_MSE_space_corr} shows the impact of the MSSC on the minimum average MSE of reconstruction by the proposed adaptation methods. For comparison, the results of the previous work \cite{Huang, Nadeem, Wu, Roth} of no inference with fixed blocklength and the existing asynchronous inference-related work with time shift adaptation only \cite{Hribar} are presented. The proposed adaptation methods achieve results close to exhaustive search at a much lower complexity. The proposed JTSBO algorithm outperforms the approach with time shift optimization only \cite{Hribar}. The inference with synchronous transmission and inference with asynchronous transmission schemes with the proposed adaptation method achieve the highest average MSE performance of reconstruction at MSSC $\bar \rho _m^{\text{spatial}}\in$ [0.03, 0.73) and $\bar \rho _m^{\text{spatial}}\in$ (0.73, 1], respectively.  At MSSC $\bar \rho _m^{\text{spatial}}=$ 1, the two schemes respectively enable an average MSE reduction of up to 50$\%$ and 63$\%$ over the no inference case \cite{Huang, Nadeem, Wu, Roth}.

	\begin{figure}[t!] 
	\centering
	{\includegraphics[height=5.5cm]{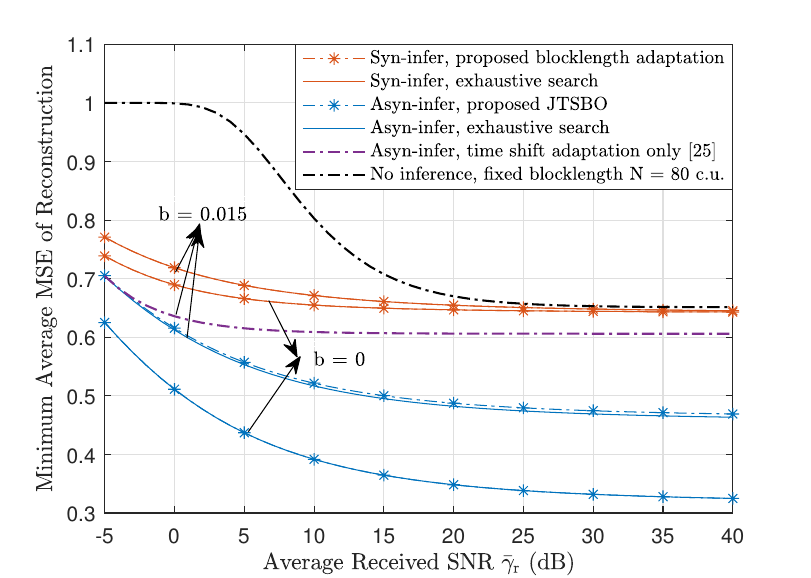}}
	\caption{Impact of average received SNR on the minimum average MSE of reconstruction by the proposed adaptation methods at transmission period $T=$ 500 ms.}
	\label{fig_Min_MSE_SNR}
\end{figure}

Fig. \ref{fig_Min_MSE_SNR} shows the impact of average received SNR on the minimum average MSE of reconstruction by the proposed adaptation methods. The minimum average MSEs decrease with a higher average received SNR, due to the decrease of average BLEP. Also, they decrease with a smaller $b$, as the spatial correlation between data increases. At $b=$ 0 and average received SNR $\bar\gamma_{\text{r}}=$ 5 dB, the inference with synchronous transmission and inference with asynchronous transmission schemes achieve an average MSE reduction of 30$\%$ and 54$\%$ over the no inference case \cite{Huang, Nadeem, Wu, Roth}, respectively. At $b=$ 0 and high average received SNR, the proposed JTSBO algorithm enables a 45$\%$ reduction in the average MSE of reconstruction over the approach with time shift optimization only \cite{Hribar}.



\section{Conclusion}

In this paper, we have presented a comprehensive analysis of the relationship between the average MSE of reconstruction and average BLEP and spatial correlation. The average MSE of reconstruction has been proved to be mono-increasing with respect to average BLEP under inference with synchronous transmission scheme, and decreasing first and then increasing with respect to average BLEP in some conditions under inference with asynchronous transmission scheme. In addition, it is mono-decreasing with respect to the MSSC under both schemes, and upper and lower bounded with respect to average BLEP and MSSC. 
Also, we have presented the preference regions of the inference with synchronous transmission and inference with asynchronous transmission schemes, by deriving the MSSC thresholds. It has been shown that the former is preferable with relatively weak MSSC, low-medium average received SNR and medium transmission period, while the latter is desirable with relatively strong MSSC, low average received SNR and long transmission period. Furthermore, blocklength and time shift adaptations have been conducted to minimize the average MSE of reconstruction. Simulation results have shown that the proposed JTSBO algorithm achieves a 45$\%$ reduction in the average MSE of reconstruction over the approach with time shift optimization only \cite{Hribar}. Compared to the no inference case \cite{Huang, Nadeem, Wu, Roth}, the inference with asynchronous transmission scheme enables an average MSE reduction of up 63$\%$, and the inference with synchronous transmission scheme enables an average MSE reduction of 50$\%$ and the performance gain is robust, even under weak MSSC and weaker than the squared temporal correlation at a transmission period length. In the future work, we will consider an event-triggered inference to further enhance the MSE performance of reconstruction.

\appendices

\section{Derivations of $\mathbb{E}[{D_{v,g}}]$ and $ \mathbb{E}[{e^{ - {\text 2}b{r_{m{l}}}}}({\text 1} - {e^{ - {\text 2}aD_{v,g}}})]$}

The expectation of ${D_{v,g}}$ can be calculated as the average of the inter-reception time over each successful transmission period, \emph{i.e.}, $\mathbb{E}[{D_{v,g}}] = \frac{{\mathbb{E}[{D_v}]}}{{\mathbb{E}[{G_v}]}}=\frac{{\frac{T}{{{\text 1} - {{\bar \varepsilon }^M}}}}}{{\sum\limits_{i = {\text 1}}^M {\frac{{{\text 1} - \bar \varepsilon }}{{{\text 1} - {{\bar \varepsilon }^M}}}} }}= \frac{T}{{M({\text 1} - \bar \varepsilon )}}$. The expectation of ${e^{ - {\text 2}b{r_{m{l}}}}}({\text 1} - {e^{ - {\text 2}aD_{v,g}}})$ can be derived by
$\mathbb{E}[{e^{ - {\text 2}b{r_{m{l}}}}}({\text 1} - {e^{ - {\text 2}aD_{v,g}}})]=\sum\limits_{n = {\text 1}}^M {\Pr \{ {l} = n\} }{e^{ - {\text 2}b{r_{mn}}}} \mathbb{E}[{\text 1} - {e^{ - {\text 2}aD_{v,g}}}|{l} = n]$, according to the law of total expectation. By mathematical induction, we obtain that given the last successful sensor ${\text S}_{l}={\text S}_n$, the inter-reception time of asynchronous transmission ${D_{v,g}}\in\{zT+ih,zT+T -(n - d)h\}$, with probabilities $\Pr \{ {D_{v,g}} =zT+ih\}  = {{\bar \varepsilon }^{zM}}{{\bar \varepsilon }^{i - {\text 1} }}({\text 1} - \bar \varepsilon )$ and $\Pr \{ {D_{v,g}} = zT+T - (n - d)h\}  = {{\bar \varepsilon }^{zM}}{{\bar \varepsilon }^{M-(n-d) - {\text 1}}}({\text 1} - \bar \varepsilon )$, where $d \in [\text1,n]$ and $i \in [\text1,M - n]$. Therefore, we have $ \mathbb{E}[{\text 1} - {e^{ - {\text 2}aD_{v,g}}}|{l} = n]= {\sum\limits_{d = {\text{1}}}^n {\mathop \sum \limits_{z = {\text{0}}}^\infty  } } {({\text{1}} - {e^{ - {\text{2}}a(zT + T - (n - d)h)}}){{\bar \varepsilon }^{zM}}{{\bar \varepsilon }^{M - (n - d) - {\text{1}}}}({\text{1}} - \bar \varepsilon )} +        \sum\limits_{i = {\text{1}}}^{M - n} {\sum\limits_{z = {\text{0}}}^\infty  ( } {\text{1}} - {e^{ - {\text{2}}a(zT + ih)}}){{\bar \varepsilon }^{zM}}{{\bar \varepsilon }^{i - {\text{1}}}}({\text{1}} - \bar \varepsilon )=\frac{{\Psi_n}}{{\text1 - {e^{-{\text{2}}ah}}\bar \varepsilon }}$. With $\Pr\{l=n \}=\frac{{\text 1}}{M}$, we obtain $ \mathbb{E}[{e^{ - {\text 2}b{r_{m{l}}}}}({\text 1} - {e^{ - {\text 2}aD_{v,g}}})]=\frac{{\sum\limits_{n = {\text{1}}}^M {{e^{ - {\text{2}}b{r_{mn}}}}\Psi_n}}}{{M(\text1 - {e^{ - {\text{2}}ah}}\bar \varepsilon )}} $.

\section{Proof of Proposition 2}

We first prove the nonotonicity of $\overline {{\text{MSE}}} _{m}^{{\text{Syn-infer}}}$ with respect to average BLEP $\bar \varepsilon$.
The first derivative of $\overline {{\text{MSE}}} _m^{{\text{Syn-infer}}}$ with respect to $\bar \varepsilon$ is given by $\frac{{\partial \overline {{\text{MSE}}} _m^{{\text{Syn-infer}}}}}{{\partial \bar \varepsilon }} =- \frac{{\sigma _{{\text{X,}}m}^{\text{2}}{\gamma _{\text{o}}}{e^{ - {\text{2}}a\tau }}({\text{1}} - {e^{ - {\text{2}}aT}})}}{{{\text{2}}aT({\gamma _{\text{o}}} + {\text{1}}){{({\text{1}} - {e^{ - {\text{2}}aT}}{{\bar \varepsilon }^M})}^{\text2}}}}\varpi$, where $\varpi  =- {\bar \varepsilon }^{-\text2}\sum\limits_{\tilde s = \text1}^M {{e^{ - \text2b{r_{m\tilde s}}}}{{\bar \varepsilon }^{\tilde s }}} [(\text1 - {e^{ - \text2aT}}{{\bar \varepsilon }^M})(\text1 - \tilde s(\text1 - \bar \varepsilon )) - M{e^{ - \text2aT}}({{\bar \varepsilon }^M} - {{\bar \varepsilon }^{M + \text1}})] \approx - (\text1 - {e^{ - \text2aT}}{{\bar \varepsilon }^M}){\bar \varepsilon }^{-\text2}\sum\limits_{\tilde s = \text1}^M {{e^{ - \text2b{r_{m\tilde s}}}}{{\bar \varepsilon }^{\tilde s}}} (\text1 - \tilde s(\text1 - \bar \varepsilon ))$ with ${{\bar \varepsilon }^M} \approx {{\bar \varepsilon }^{M + \text1}}$. Since $\sum\limits_{\tilde s= {\text{1}}}^{M} {{e^{ - {\text{2}}b{r_{m\tilde s}}}}} {{\bar \varepsilon }^{\tilde s}}(\text1 - \tilde s(\text1 - \bar \varepsilon ))$ is an increasing function of $\bar \varepsilon $ and is equal to zero at $\bar \varepsilon= $ 0, we have $\varpi<0$  for $ \bar \varepsilon\in$ (0, 1),  and hence $\frac{{\partial \overline {{\text{MSE}}} _m^{{\text{Syn-infer}}}}}{{\partial \bar \varepsilon }}>$ 0 holds. 

Next, we prove the nonotonicity of $\overline {{\text{MSE}}} _{m}^{{\text{Asyn-infer}}}$ with respect to average BLEP $\bar \varepsilon$. We consider two cases of time shift $h=\frac{T}{M}$ and $h\ne\frac{T}{M}$. In the special case of $h=\frac{T}{M}$, the first derivative of $\overline {{\text{MSE}}} _m^{{\text{Asyn-infer}}}$ with respect to $\bar \varepsilon$ is given by $\frac{{\partial  \overline {{\text{MSE}}} _m^{{\text{Asyn-infer}}}}}{{\partial \bar \varepsilon }}= \frac{{\sigma _{{\text{X}},m}^{\text{2}}{\gamma _{\text{o}}}{e^{ - {\text{2}}a\tau }}{{({\text{1}} - {e^{ - {\text{2}}ah}})}^{\text2}}\sum\limits_{n = {\text{1}}}^M {{e^{ - {\text{2}}b{r_{mn}}}}} }}{{{\text{2}}aT({\gamma _{\text{o}}} + {\text{1}}){{({\text{1}} - {e^{ - {\text{2}}ah}}\bar \varepsilon )}^{\text2}}}}>$ 0. Therefore, $\overline {{\text{MSE}}} _{m}^{{\text{Asyn-infer}}}$ is mono-increasing with respect to $\bar \varepsilon$ with  $h=\frac{T}{M}$. In the case of $h\ne\frac{T}{M}$, the first derivative of $\overline {{\text{MSE}}} _m^{{\text{Asyn-infer}}}$ with respect to $\bar \varepsilon$ is given by $\frac{{\partial \overline {{\text{MSE}}} _m^{{\text{Asyn-infer}}}}}{{\partial \bar \varepsilon }}= - \frac{{\sigma _{{\text{X}},m}^{\text{2}}{\gamma _{\text{o}}}{e^{ - {\text{2}}a\tau }}({\text{1}} - \bar \varepsilon )}}{{{\text{2}}aT({\gamma _{\text{o}}} + {\text{1}}){{({\text{1}} - {e^{ - {\text{2}}ah}}\bar \varepsilon )}^{\text2}}}}  [({\text{1}} - {e^{ - {\text{2}}ah}}\bar \varepsilon )\sum\limits_{n = {\text{1}}}^M {{e^{ - {\text{2}}b{r_{mn}}}}\frac{{{\Psi _n}}}{{\partial \bar \varepsilon }}}  - {e^{ - {\text{2}}ah}}(\text1 - {e^{ - {\text{2}}ah}})\sum\limits_{n = {\text{1}}}^M {{e^{ - {\text{2}}b{r_{mn}}}}{\Psi _n}} ] $, where $\frac{{\partial {\Psi _n}}}{{\partial \bar \varepsilon }} = \frac{{{e^{ - {\text{2}}ah(\text1-n)}}({e^{ - {\text{2}}ahM}} - {e^{ - {\text{2}}aT}}){{\bar \varepsilon }^{M - n -\text 1}}}}{{{{(\text1 - {e^{ - {\text{2}}aT}}{{\bar \varepsilon }^M})}^{\text2}}}}[(M - n)(\text1- \bar \varepsilon )$ $ - \bar \varepsilon  + {e^{ - {\text{2}}aT}}{{\bar \varepsilon }^M}(n(\text1 - \bar \varepsilon ) + \bar \varepsilon )]$. When $\bar \varepsilon  \to \text1^-$, we have $\frac{{\partial \overline {{\text{MSE}}} _m^{{\text{Asyn-infer}}}}}{{\partial \bar \varepsilon }} \to  \text0^+$, \emph{i.e.}, $\overline {{\text{MSE}}} _m^{{\text{Asyn-infer}}}$ increases with $\bar \varepsilon$ at very high $\bar \varepsilon$. When $\bar \varepsilon  \to \text0^+$, we obtain $\frac{{\partial \overline {{\text{MSE}}} _m^{{\text{Asyn-infer}}}}}{{\partial \bar \varepsilon }} \to   \frac{{\sigma _{{\text{X}},m}^{\text{2}}{\gamma _{\text{o}}}{e^{ - {\text{2}}a\tau }}{e^{ - {\text{2}}ah}}\Theta }}{{{\text{2}}aT({\gamma _{\text{o}}} + {\text{1}}){{(\text1 - {e^{ - {\text{2}}ah}})}^{\text2}}}}$, where $\Theta= \sum\limits_{n = \text1}^M {{e^{ - \text2b{r_{mn}}}}}  - {e^{ - \text2ah}}{(\text1 - {e^{ - \text2ah}})^{ - \text2}}(\text1 - {e^{ - \text2aT}}{e^{\text2ahM}})$ $\left( {{e^{ - \text2b{r_{m(M - \text1)}}}} - (\text1 - {e^{ - \text2ah}}){e^{ -\text 2b{r_{mM}}}}} \right) =(M-\text1)(\bar \rho _m^{{\text{spatial}}}-\Upsilon)$. Therefore, we get when $\bar \varepsilon  \to \text0^+$, $\frac{{\partial \overline {{\text{MSE}}} _m^{{\text{Asyn-infer}}}}}{{\partial \bar \varepsilon }}>$ 0 if $\bar \rho _m^{{\text{spatial}}} > \Upsilon $, and $\frac{{\partial \overline {{\text{MSE}}} _m^{{\text{Asyn-infer}}}}}{{\partial \bar \varepsilon }}<$ 0 if $\bar \rho _m^{{\text{spatial}}} < \Upsilon $. Since $\overline {{\text{MSE}}} _{m}^{{\text{Asyn-infer}}}$ is convex with respect to $\bar \varepsilon$, which is validated by simulation and shown in Fig. \ref{3d_asyn} (a) and (b). Together with  $\frac{{\partial \overline {{\text{MSE}}} _m^{{\text{Asyn-infer}}}}}{{\partial \bar \varepsilon }} \to  \text0^+$ when $\bar \varepsilon  \to \text1^-$. It is obtained that $\overline {{\text{MSE}}} _{m}^{{\text{Asyn-infer}}}$ is mono-increasing with respect to $\bar \varepsilon$ with $\bar \rho _m^{{\text{spatial}}} > \Upsilon $, and decreasing first and then increasing  with respect to $\bar \varepsilon$ with $\bar \rho _m^{{\text{spatial}}} < \Upsilon $, in the case of $h\ne\frac{T}{M}$. Hence, Proposition 2 is guaranteed.

\section{Proof of Proposition 5}

The second derivative of $\overline {{\text{MSE}}} _m^{{\text{Syn-infer}}}$ with respect to blocklength $N$ is given by $\frac{{{\partial ^{\text 2}}\overline {{\text{MSE}}} _m^{{\text{Syn-infer}}}}}{{\partial {N^{\text 2}}}}=-\frac{{\sigma _{{\text{X,}}m}^{\text{2}}{\gamma _{\text{o}}}{e^{ - {\text{2}}a\tau }}({\text{1}} - {e^{ - {\text{2}}aT}})}}{{{\text{2}}aT({\gamma _{\text{o}}} + {\text{1}})}}[{{{-({\text{2}}a{T_{\text{s}}})^{\text2}}}\hbar  - \text4a{T_{\text{s}}}\frac{{\partial \hbar }}{{\partial \bar \varepsilon }}\frac{{\partial \bar \varepsilon }}{{\partial N}}}{+ \frac{{{\partial ^{\text2}}\hbar }}{{\partial {{\bar \varepsilon }^{\text2}}}}{{(\frac{{\partial \bar \varepsilon }}{{\partial N}})^{\text2}}}}+{\frac{{\partial \hbar }}{{\partial \bar \varepsilon }}}  {\frac{{{\partial ^{\text2}}\bar \varepsilon }}{{\partial {N^{\text2}}}}}]$, where $\hbar  = \frac{{{\text{1}} - \bar \varepsilon }}{{{\text{1}} - {e^{ - {\text{2}}aT}}{{\bar \varepsilon }^M}}}\sum\limits_{\tilde s = {\text{1}}}^M {{e^{ - {\text{2}}b{r_{m\tilde s}}}}{{\bar \varepsilon }^{\tilde s - {\text{1}}}}}$. We first prove $\frac{{\partial \hbar }}{{\partial \bar \varepsilon }}<$ 0 and $\frac{{{\partial ^{\text2}}\hbar }}{{\partial {{\bar \varepsilon }^{\text2}}}} <$ 0. They are respectively expressed as $\frac{{\partial \hbar }}{{\partial \bar \varepsilon }}= \frac{\varpi }{{{{({\text{1}} - {e^{ - {\text{2}}aT}}{{\bar \varepsilon }^M})}^{\text{2}}}}}$ and $\frac{{{\partial ^{\text2}}\hbar }}{{\partial {{\bar \varepsilon }^{\text2}}}} = \frac{{\text2M{e^{ - {\text{2}}aT}}{{\bar \varepsilon }^{M -\text 1}}\varpi  + ({\text{1}} - {e^{ - {\text{2}}aT}}{{\bar \varepsilon }^M})\frac{{\partial \varpi }}{{\partial \bar \varepsilon }}}}{{{{({\rm\text{1}} - {e^{ - {\text{2}}aT}}{{\bar \varepsilon }^M})}^{\text3}}}}$, where  $\frac{{\partial \varpi }}{{\partial \bar \varepsilon }} =-{{\bar \varepsilon }^{-\text3}}  \sum\limits_{\tilde s = \text1}^M {{e^{ - \text2b{r_{m\tilde s}}}}{{\bar \varepsilon }^{\tilde s}}} [(\text1 - {e^{ - \text2aT}}{{\bar \varepsilon }^M})(\tilde s -\text 1)(\text2 - \tilde s(\text1 - \bar \varepsilon )) + M\tilde s{e^{ - \text2aT}}{{\bar \varepsilon }^M}(\text1 - \bar \varepsilon ) + M{e^{ - \text2aT}}(M + \tilde s - \text1)({{\bar \varepsilon }^M} - {{\bar \varepsilon }^{M + \text1}}) + M{e^{ - \text2aT}}({{\bar \varepsilon }^{M + \text1}} - {{\bar \varepsilon }^M})]$. We have proved $\varpi<$ 0 in the proof of Proposition 2, and hence $\frac{{\partial \hbar }}{{\partial \bar \varepsilon }}<$ 0 holds. Furthermore, using ${{\bar \varepsilon }^M} \approx {{\bar \varepsilon }^{M + \text1}}$ on the last two terms of the sum function in the expression of $\frac{{\partial \varpi }}{{\partial \bar \varepsilon }}$ yields $\frac{{\partial \varpi }}{{\partial \bar \varepsilon }} \approx -\bar \varepsilon^{-\text3}\sum\limits_{\tilde s = {\text{1}}}^M {{e^{ - {\text{2}}b{r_{m\tilde s}}}}{{\bar \varepsilon }^{\tilde s }}} [({\text{1}} - {e^{ - {\text{2}}aT}}{{\bar \varepsilon }^M})(\tilde s - \text1)(\text2 - \tilde s(\text1 - \bar \varepsilon )) + M\tilde s{e^{ - {\text{2}}aT}}{{\bar \varepsilon }^M}(\text1 - \bar \varepsilon )]$. Since $\sum\limits_{\tilde s = {\text{1}}}^M {{e^{ - {\text{2}}b{r_{m\tilde s}}}}{{\bar \varepsilon }^{\tilde s}}} (\tilde s - \text1)(\text2 - \tilde s(\text1 - \bar \varepsilon ))$ is an increasing function of $ \bar \varepsilon$ and is equal to zero at $ \bar \varepsilon=$ 0, we have $\frac{{\partial \varpi }}{{\partial \bar \varepsilon }} < -M{e^{ - {\text{2}}aT}}{{\bar \varepsilon }^{M-\text3}}(\text1 - \bar \varepsilon )\sum\limits_{\tilde s = {\text{1}}}^M {{e^{ - {\text{2}}b{r_{m\tilde s}}}}{{\bar \varepsilon }^{\tilde s}}} \tilde s<$ 0 for $ \bar \varepsilon\in$ (0, 1). Therefore, $\frac{{{\partial ^{\text2}}\hbar }}{{\partial {{\bar \varepsilon }^{\text2}}}}<$ 0 holds. 

Next, we prove $\frac{{\partial \bar \varepsilon }}{{\partial N}}<$ 0 and $\frac{{{\partial ^{\text2}}\bar \varepsilon }}{{\partial {N^{\text2}}}}>$ 0.  With ${{{\bar \varepsilon }}} \approx {\text 1} - {e^{ - \frac{{\text 1}}{{{{\bar \gamma }_{{\text r}}}}}(\eta  - \sqrt {\pi L} /N)}}$ , we have $\frac{{\partial {{\bar \varepsilon }}}}{{\partial N}} \approx \frac{{(\sqrt {\pi L}  - L{e^{L/N}}){e^{ - \frac{{\text 1}}{{{{\bar \gamma }_{{\text r}}}}}(\eta  - \sqrt {\pi L} /N)}}}}{{{{\bar \gamma }_{{\text r}}}{N^{\text 2}}}}$, which is smaller than zero with $L \ge \pi$ \cite{Yu}. Based on the first-order Taylor approximation of ${e^{ - \frac{{\text{1}}}{{{{\bar \gamma }_{\text{r}}}}}(\eta  - \sqrt {\pi L} /N)}} \approx \text1 - \frac{{\text{1}}}{{{{\bar \gamma }_{\text{r}}}}}(\eta  - \sqrt {\pi L} /N)$, we have $\frac{{{\partial ^{\text 2}}{{\bar \varepsilon }}}}{{\partial {N^{\text 2}}}} \approx \frac{{({L^{\text2}} + \text2LN){e^{L/N}} - \text2N\sqrt {\pi L} }}{{{{\bar \gamma }_{\text{r}}}{N^{\text4}}}}>$ 0 with $L \ge \pi$ \cite{Yu}. Therefore,  $\frac{{{\partial ^{\text2}}\overline {{\text{MSE}}} _m^{{\text{Syn-infer}}}}}{{\partial {N^{\text2}}}}>$ 0 is guaranteed. With constraint (C1) being an affine set, Proposition 5 is proved.

\section{Proof of Proposition 6}
The second derivative of $\overline {{\text{MSE}}} _m^{{\text{Asyn-infer}}}$ with respect to time shift $h$ is given by $\frac{{{\partial ^{\text 2}}\overline {{\text{MSE}}} _m^{{\text{Asyn-infer }}}}}{{\partial {h^{\text2}}}} = \frac{{{\text2}a\sigma _{{\text{X}},m}^{\text{2}}{\gamma _{\text{o}}}{e^{ - {\text{2}}a\tau }}({\text{1}} - \bar \varepsilon )}}{{T({\gamma _{\text{o}}} + {\text{1}}){{({\text{1}} - {e^{ - {\text{2}}ah}}\bar \varepsilon )}^{\text3}}}}\sum\limits_{n = {\text{1}}}^M {{e^{ - {\text{2}}b{r_{mn}}}}} ({\chi  _{\text1}} + {\chi_{\text2}} + {\chi_{\text3}})$, where ${\chi_\text1} = {e^{ - {\text{2}}ah}}(\text1 - \bar \varepsilon )(\text1+ \bar \varepsilon {e^{ - {\text{2}}ah}})$, ${\chi_{\text2}} = \vartheta {e^{ - {\text2}a(T + h(\text1 - n))}}[{(n - \text1)^{\text2}} + (\text1 + {\text2}n - {n^{\text2}}({\text2} - {e^{ - {\text{2}}ah}}\bar \varepsilon )){e^{ - {\text{2}}ah}}\bar \varepsilon ]$, ${\chi_{\text3}} = \vartheta {e^{ - {\text{2}}ah(M - n + \text1)}}[ - {(M - n + \text1)^{\text2}} - (\text1 - {\text2}(M - n) - {(M - n)^{\text2}}({\text2} - {e^{ - {\text{2}}ah}}\bar \varepsilon )){e^{ - {\text{2}}ah}}\bar \varepsilon ]$ and $\vartheta  = \frac{{{{\bar \varepsilon }^{M - n}}({\text{\text1}} - \bar \varepsilon )}}{{{\text{(\text1}} - {e^{ - {\text{2}}aT}}{{\bar \varepsilon }^M})}}$. We only need to prove the sum function on the right-hand side is greater than zero. With $\chi_{\text1}>$ 0, we prove ${\sum\limits_{n = {\text{\text1}}}^M {{e^{ - {\text{2}}b{r_{mn}}}}} }(\chi_{\text2}+\chi_{\text3})>$ 0. With $T\ge Mh$, we have $\chi_{\text2} + {\chi_{\text3}} \ge \vartheta {e^{ - {\text2}a(T + h(\text1 - n))}}M(\text1 - {e^{ - {\text{2}}ah}}\bar \varepsilon )[({\text2}n - M)(\text1 - {e^{ - {\text{2}}ah}}\bar \varepsilon ) - {\text2}]$, which can be approximated as $  \vartheta {e^{ - {\text2}a(T + h(\text1 - n))}}M{(\text1 - {e^{ - {\text{2}}ah}}\bar \varepsilon )^{\text2}}({\text2}n - M)$, as the value of (${\chi_{\text2}} + {\chi_{\text3}}$) is dominated by $\vartheta {e^{ - {\text2}a(T + h(\text1 - n))}}$. Then, we obtain $\sum\limits_{n = {\text{\text1}}}^M {{e^{ - {\text{2}}b{r_{mn}}}}} ({\chi_{\text2}} + {\chi_{\text3}}) \ge \varsigma {M^{\text2}}{(\text1 - {e^{ - {\text{2}}ah}}\bar \varepsilon )^{\text2}}>$ 0, with $\varsigma  = \mathop {\min }\limits_{n \in [\text1,M]} {e^{ - {\text{2}}b{r_{mn}}}}\vartheta {e^{ - {\text2}a(T + h(\text1 - n))}}$. Therefore, $\frac{{{\partial ^{\text 2}}\overline {{\text{MSE}}} _m^{{\text{Asyn-infer }}}}}{{\partial {h^{\text2}}}}>$ 0 is guaranteed.

Given time shift $h=\frac{T}{M}$, the second derivative of $\overline {{\text{MSE}}} _m^{{\text{Asyn-infer}}}$ with respect to blocklength $N$ is given by $\frac{{{\partial ^{\text2}}\overline {{\text{MSE}}} _m^{{\text{Asyn-infer}}}}}{{\partial {N^{\text2}}}} =\frac{{\sigma _{{\text{X}},m}^{\text{2}}{\gamma _{\text{o}}}({\text{1}} - {e^{ - {\text{2}}ah}})\sum\limits_{n = {\text{1}}}^M {{e^{ - {\text{2}}b{r_{mn}}}}}}}{{{\text{2}}aT({\gamma _{\text{o}}} + {\text{1}}){{({\text{1}} - {e^{ - {\text{2}}ah}}\bar \varepsilon )}^{\text3}}}} [(\text1 - {e^{ - {\text{2}}ah}})({\text{1}} - {e^{ - {\text{2}}ah}}\bar \varepsilon )\frac{{{\partial ^{\text2}}\bar \varepsilon }}{{\partial {N^{\text2}}}} + \text2{e^{ - {\text{2}}ah}}(\text1 - {e^{ - {\text{2}}ah}}){(\frac{{\partial \bar \varepsilon }}{{\partial N}})^{\text2}} - \text2a{T_{\text{s}}}(\text1 + {e^{ - {\text{2}}ah}}\bar \varepsilon  - {e^{ - {\text{2}}ah}})\frac{{\partial \bar \varepsilon }}{{\partial N}}] $. We have proved ${\frac{{\partial \bar \varepsilon }}{{\partial N}}}<$ 0 and ${\frac{{{\partial ^{\text2}}\bar \varepsilon }}{{\partial {N^{\text2}}}}}>$ 0 in the proof of Proposition 5. Therefore, $\frac{{{\partial ^{\text2}}\overline {{\text{MSE}}} _m^{{\text{Asyn-infer}}}}}{{\partial {N^{\text2}}}}>$ 0 is guaranteed.

\end{document}